\documentclass[11pt,a4paper]{article}

\pdfoutput=1

\usepackage{amssymb}
\usepackage{amsmath}
\usepackage{amsfonts}
\usepackage{graphicx}
\usepackage{booktabs}
\usepackage{color}
\usepackage{textcomp}
\usepackage{multirow}
\usepackage{bm}
\usepackage{caption}
\usepackage{subcaption}
\usepackage{tikz}
\usepackage[export]{adjustbox}
\usepackage{longtable}
\usepackage{colortbl}
\usepackage{rotating}
\usepackage{jheppub}
\usepackage[utf8]{inputenc}
\usepackage{braket}
\usepackage{hyperref}

\usepackage{epstopdf}
\usepackage[labelfont=bf]{caption}
\usepackage[section]{placeins}
\usepackage{bbold}

\definecolor{jlab_red}{RGB}{192,39,45}
\definecolor{jlab_orange}{RGB}{249,102,0}
\definecolor{jlab_blue}{RGB}{47,122,121}
\definecolor{jlab_green}{RGB}{65,125,10}

\definecolor{swave}{HTML}{B23B3B}
\definecolor{pwave}{HTML}{424874}
\definecolor{dstarpi}{HTML}{A15017} 

\newcommand{\cm}{\ensuremath{\mathsf{cm}}}
\newcommand{\nn}{\mathfrak{n}}

\renewcommand{\star}{\ensuremath{\ast}}

\hypersetup{%
pdftitle = {Isospin-1/2 D-pi scattering and the lightest D_0-star resonance from lattice QCD},
pdfauthor = {Hadron Spectrum Collaboration},
colorlinks = {true},
filecolor = {black},
linkcolor = {jlab_blue},
menucolor = {black},
citecolor = {jlab_green},
urlcolor = {jlab_green},
}{}

\begin{document}

\title{\boldmath Isospin-1/2 $D\pi$ scattering and the lightest $D_0^\ast$ resonance from lattice QCD}

\author[a]{Luke Gayer,}
\author[a]{Nicolas Lang,}
\author[a]{Sin\'ead M. Ryan,}
\author[a]{David Tims,\footnote{deceased}}
\author[b]{Christopher~E.~Thomas,}
\author[b]{David J. Wilson}
\author{\\ \textnormal{(for the Hadron Spectrum Collaboration)}}

\affiliation[a]{School of Mathematics and Hamilton Mathematics Institute, Trinity College, Dublin 2, Ireland}
\affiliation[b]{Department of Applied Mathematics and Theoretical Physics, Centre for Mathematical Sciences,\\ University of Cambridge, Wilberforce Road, Cambridge, CB3 0WA, UK}

\emailAdd{lgayer@tcd.ie}
\emailAdd{nicolas.lang@maths.tcd.ie}
\emailAdd{ryan@maths.tcd.ie}
\emailAdd{c.e.thomas@damtp.cam.ac.uk}
\emailAdd{d.j.wilson@damtp.cam.ac.uk}

\abstract{Isospin-1/2 $D\pi$ scattering amplitudes are computed using lattice QCD, working in a single volume of approximately $(3.6\; \mathrm{fm})^3$ and with a light quark mass corresponding to $m_\pi\approx239$ MeV. The spectrum of the elastic $D\pi$ energy region is computed yielding 20 energy levels. Using the L\"uscher finite-volume quantisation condition, these energies are translated into constraints on the infinite-volume scattering amplitudes and hence enable us to map out the energy dependence of elastic $D\pi$ scattering. By analytically continuing a range of scattering amplitudes, a $D_0^\ast$ resonance pole is consistently found strongly coupled to the $S$-wave $D\pi$ channel, with a mass $m\approx 2200$~MeV and a width $\Gamma\approx400$ MeV. Combined with earlier work investigating the $D_{s0}^\ast$, and $D_0^\ast$ with heavier light quarks, similar couplings between each of these scalar states and their relevant meson-meson scattering channels are determined. The mass of the $D_0^\ast$ is consistently found well below that of the $D_{s0}^\ast$, in contrast to the currently reported experimental result.}

\maketitle

\section{Introduction}
\label{sec:intro}

Since their experimental discoveries in 2003, the lightest scalar charm-light $D_{0}^\star$ and charm-strange $D_{s0}^\star$ mesons have stimulated much theoretical activity. Within the quark-model they have a common $P$-wave $q\bar{q}$ construction and the origin of much of the mass difference is due to the differing light and strange quark masses. This results in predictions of two similar states with the mass of the charm-light state below the charm-strange state. However, the states have been observed experimentally at similar masses with vastly different widths, the $D_0^\star$ seen as a broad feature in $D\pi$ amplitudes, in contrast with a narrow $D_{s0}^\star$ found below $DK$ threshold.

The lightest experimentally observed scalar $D_0^\star$ resonance was first reported by the Belle~\cite{Abe:2003zm} and FOCUS~\cite{Link:2003bd} experiments with a mass in the range 2300-2400~MeV. Later results from BABAR~\cite{Aubert:2009wg} and a pair of studies by LHCb~\cite{Aaij:2015kqa,Aaij:2015sqa} also found a broad resonance at a similar mass, not so different from quark potential models that predict a $D_0^\star$ around 2400 MeV~\cite{Godfrey:1985xj}. The $D_{s0}^\star(2317)$ was observed as a narrow peak in isospin-breaking $D_s\pi$ final states, with a mass below $DK$ threshold. In contrast to the $D_0^\star$, the $D_{s0}^\star$  is reported some 200~MeV below typical quark potential model predictions~\cite{Zyla:2020zbs}, perhaps even below the $D_0^\star$. The vastly differing widths, and how these affect the masses, call for a deeper theoretical analysis of these states and the light quark mass dependence in the open charm sector~\cite{Eichten:2019may}. 

The limitations of quark potential models in scalar systems are understood. In particular, since decay channels made of pairs of pseudoscalars open with no angular momentum barrier they can have a significant effect, both when nearby but kinematically closed and when a strong coupling to a decay channel produces a large width. Many puzzling hadronic resonances arise close to thresholds, from the $D_0^\star$ and $\chi_{c1}(3872)$, to more recent observations in exotic flavour such as $D^-K^+$ (with flavour content $\bar{c}\bar{s}du$)~\cite{Aaij:2020hon,Aaij:2020ypa} and $J/\psi\;J/\psi$ final states~\cite{Aaij:2020fnh}, and so a model-independent approach is required to explore fully the QCD dynamics. 

While experimental production of these hadrons proceeds through heavy meson decays, the simplest theoretical perspective is to observe them as part of scattering processes such as $D\pi\to D\pi$. Lattice QCD is a first-principles approach that has been applied successfully to a range of scattering processes. In the light meson sector properties of simple narrow resonances such as the $\rho$ seen in $P$-wave $\pi\pi\to\pi\pi$ scattering have been computed~\cite{Feng:2010es,Dudek:2012xn,Wilson:2015dqa,Bulava:2016mks,Alexandrou:2017mpi,Andersen:2018mau,Werner:2019hxc,Erben:2019nmx}. Scalar resonances have also been studied in elastic and coupled-channel systems including the $\sigma$~\cite{Briceno:2016mjc}, the $f_0$~\cite{Briceno:2017qmb} and the $a_0$~\cite{Dudek:2016cru}. More recently, states involving spinning scattering hadrons have been determined, such as the $b_1$~\cite{Woss:2019hse}, an exotic $\pi_1$~\cite{Woss:2020ayi}, and several $J^{PC}=J^{--}$ resonances~\cite{Johnson:2020ilc}.

In Ref.~\cite{Moir:2016srx} $D\pi$,~$D\eta$,~$D_s\bar{K}$ scattering amplitudes were computed from 47 energy levels in the first coupled-channel calculation involving charm quarks using lattice QCD. A near-threshold scalar $D_0^\star$ bound-state was identified albeit with a heavier-than-physical light quark mass corresponding to $m_\pi=391$~MeV. In this study, we compute $D\pi$ scattering amplitudes with $m_\pi=239$~MeV to investigate how the lightest $D_0^\star$ evolves with the light quark mass. We have previously investigated several other channels at these two light quark masses, beginning with the $\rho$~\cite{Dudek:2012xn,Wilson:2015dqa}, followed by the $\sigma$~\cite{Briceno:2016mjc} and $\pi K$~\cite{Wilson:2014cna,Wilson:2019wfr}. Crucially for the present study, $DK$~\cite{Cheung:2020mql} was also determined at both masses and those results combined with Ref.~\cite{Moir:2016srx} allowed for a first-principles comparison of the $D_{s0}^\star$ and $D_0^\star$ at $m_\pi=391$~MeV. In this case, the $D_0^\star$ was found significantly below the $D_{s0}^\star$. In the present study, we complete the quartet of calculations to better understand both systems.

Charm-light systems are featured in several other studies using lattice QCD~\cite{Liu:2012zya, Mohler:2012na, Moir:2013ub, Bali:2015lka, Kalinowski:2015bwa, Cichy:2016bci, Cheung:2016bym}. Other than our earlier work~\cite{Moir:2016srx}, only Ref.~\cite{Mohler:2012na} has studied $I=1/2$ $D\pi$ scattering. The latter calculation used a small volume with no dynamical strange quarks and obtained two energy levels in the elastic scattering region. By assuming the presence of a pole and that a Breit-Wigner parameterisation describes the two energies, a resonance was found with a mass around 2400 MeV, but these assumptions could not be tested with their data.  

The paper is organised as follows. Section~\ref{sec:calc} gives the lattice parameters and calculation details. In section~\ref{sec:spectra} the finite volume spectra are presented and the scattering amplitudes subsequently determined using L\"uscher's formulation are described in section~\ref{sec:scattering}. The location of poles and their couplings are discussed in section~\ref{sec:poles}. An interpretation and an analysis of the light quark mass dependence are given in section~\ref{sec:interpretation}. A summary of this work and an outlook are given in section~\ref{sec:summary}.

\section{Calculation Details}
\label{sec:calc}

Lattice QCD is a first principles approach to QCD. Working in a finite cubic volume with a finite lattice spacing, correlation functions are computed by numerically sampling the QCD path integral in Euclidean spacetime. In a finite spatial volume, momentum is quantised and thus only discrete spectra of scattering energies are accessible. However, the mapping between the discrete spectrum obtained in a finite volume and the infinite volume scattering amplitudes
is known for two-hadron systems~\cite{Luscher:1986pf} and recent developments are reviewed in Ref.~\cite{Briceno:2017max}.

The results presented here are computed using a single lattice with $(L/a_s)^3\times(T/a_t)=32^3\times 256$, where $L^3$ is the spatial volume and $T$ is the temporal extent, with a periodic boundary condition in space and an antiperiodic boundary condition in time. An anisotropic lattice formulation is utilised in which the temporal lattice spacing, $a_t$, is finer than the spatial lattice spacing, $a_s$, and $\xi \equiv a_s/a_t \approx 3.5$. To quote results in physical units, we compare the calculated $\Omega$ baryon mass to its physical value yielding $a_t^{-1}=6.079$ GeV and a spatial lattice spacing $a_s = 0.11$ fm~\cite{Wilson:2019wfr}, corresponding to a physical spatial volume of $(3.6\; \mathrm{fm})^3$. Working with only a single lattice spacing, it is not possible to quantify the lattice spacing dependence of the results. Refs.~\cite{Liu:2012ze,Moir:2013ub,Cheung:2016bym} discuss some of these discretisation effects with charm quarks on these lattices, and Ref.~\cite{Andersen:2018mau} investigates the lattice spacing dependence in a scattering calculation of the $\rho$ resonance on other lattices and finds a relatively small effect. The lattice spacing dependence will be a topic of future studies of these systems.

A tree-level Symanzik-improved anisotropic action is used to describe the gauge sector while for the fermion sector a tree-level, tadpole-improved Sheikholeslami-Wohlert action with $N_f=2+1$ dynamical quark flavours and stout-smeared spatial gauge fields is used~\cite{Morningstar:2003gk}. This calculation uses 484 gauge configurations. The light quarks on this ensemble correspond to $m_\pi = 239$ MeV while the heavier dynamical quark is tuned to 
approximate the strange quark mass. The valence charm quark is simulated using the same action as for the light and strange quarks with mass and anisotropy parameters tuned to simultaneously reproduce the physical $\eta_c$ mass and the pion anisotropy, determined from the dispersion relations~\cite{Liu:2012ze,Cheung:2016bym}.

\subsection{Computing the spectrum}

To extract many states over a wide energy region, we follow the procedure used in Ref.~\cite{Dudek:2012xn}. We compute matrices of correlation functions
\begin{equation}
C_{ij}(t) = \braket{0 | \mathcal{O}_i(t) \mathcal{O}_j^{\dagger}(0) | 0}, 
\end{equation}
from large bases of interpolating operators $\mathcal{O}_i$ that carry the quantum numbers of the $I=1/2$ $D \pi$ system. The finite-volume eigenstates can then be obtained by solving a generalised eigenvalue problem~\cite{Michael:1985ne,Luscher:1990ck} of the form
\begin{equation}
C_{ij}(t) v_j^{(\nn)} = \lambda_\nn(t, t_0) C_{ij}(t_0) v_j^{(\nn)},   
\end{equation}
where the correlation matrix $C_{ij}$ is diagonalised timeslice-by-timeslice to obtain the eigenvectors $v^{\nn}$, and the generalised eigenvalues, or principal correlators, $\lambda_\nn(t,t_0)$. The implementation adopted here is outlined in Refs.~\cite{Dudek:2009qf,Dudek:2010wm}. The energy spectrum is extracted from an analysis of the time dependence of the principal correlators. To account for possible excited state contaminations we fit a sum of two exponentials,
\begin{equation}
\lambda_\nn (t,t_0) = (1-A_\nn) e^{-E_\nn(t-t_0)} + A_\nn e^{-E_\nn'(t-t_0)}.
\end{equation}
The parameter $E_\nn$ yields the energy of the state, while $A_\nn$ and $E_\nn^\prime$ are only used to stabilise the fit and play no further role in the subsequent analysis. A linear combination that best interpolates a given state $\nn$ can then be constructed from the eigenvector $v^\nn$ through $\Omega_\nn^{\dagger} = \sum_i v_i^{(\nn)} \mathcal{O}_i^{\dagger}$.

The basis of operators in each irrep is constructed to achieve a good overlap with the states we investigate. We use quark bilinears of the form $\bar{\psi}\Gamma D ...\psi$ where $\Gamma$ represents a product of $\gamma$-matrices, $D$ is a gauge-covariant derivative and the ellipsis indicates that up to three derivatives are used. The inclusion of derivatives allows for the construction of operators with good overlap onto higher angular momenta states, more than typically obtained using only $\gamma$-matrices. Further details can be found in Ref.~\cite{Dudek:2010wm,Thomas:2011rh}. 

The operator bases also include meson-meson-like operators which take the form $\sum_{\vec{p_1} + \vec{p_2} = \vec{P}} \mathcal{C}(\vec{p_1}, \vec{p_2}) \Omega_{M_1}^{\dagger}(\vec{p_1}) \Omega_{M_2}^{\dagger}(\vec{p_2})$, where $\Omega_{M_i}^{\dagger}(\vec{p_i})$ is a variationally-optimised operator that interpolates meson $M_i$ with lattice momentum $\vec{p}_i$. These operators 
are constructed for each momentum from eigenvectors $v^\nn$ determined in variational analyses of 
$\pi$, $K$, $\eta$, $D$, $D_s$ and $D^\star$ mesons. Further details of the construction of these operators are given in Refs.~\cite{Thomas:2011rh,Dudek:2012gj}.

The finite cubic volume breaks rotational symmetry so that eigenstates cannot be labelled by representations $J$ of the orthogonal group. Instead they are categorised by the irreducible representations (irreps) of the cubic group $O_h$ when at rest~\cite{Johnson:1982yq}, or by the little group $LG(\vec{P})$ at non-zero momentum~\cite{Moore:2005dw}. The little groups correspond to crystallographic point groups with rotation and reflection symmetry determined by the direction of the momentum. The interpolating operators are projected into irreps of the corresponding symmetry group. The subduction of continuum $J^P$ into lattice irreps relevant to this calculation is summarised in table~\ref{table:pw_irreps}. 

The operators used in constructing the correlation matrices are summarised in appendix~\ref{app:sec:ops}. We use several constructions resembling a ``single-meson'' in each irrep, and we use the non-interacting energies to guide the choice of appropriate ``meson-meson'' constructions. Non-interacting meson-meson energies are given by $E=\sqrt{m_1^2+|\vec{p_1}|^2} + \sqrt{m_2^2+|\vec{p_2}|^2}$ for the scattering of mesons 1 and 2. We include all meson-meson constructions corresponding to a non-interacting level in the energy region below the $D\pi\pi$ threshold as well as several additional operators that are only expected to produce levels at higher energies to help ensure a reliable spectrum is obtained.

\begin{table}
\begin{center}
\begin{tabular}{cc|l|l|l}
\multirow{2}{*}{$\vec{P}$} & {Irrep}  & $J^P$ $(\vec{P}=\vec{0})$  & $D\pi$ $J^P_{[N]}$ & $D^\ast\pi$ $J^P_{[N]}$\\
                           & $\Lambda$ & $|\lambda|^{(\tilde\eta)}$ $(\vec{P}\ne\vec{0})$  &  & \\
\hline
\hline
\multirow{3}{*}{$[000]$}  & 
    $A_1^+$  & $0^+$, $4^+$ &  $\bm 0^+$, ... & ...\\
  & $T_1^-$  & $1^-$, $3^-$ &  $\bm 1^-$, ... & ... \\
  & $E^+$    & $2^+$, $4^+$ &  $\bm 2^+$, ... & ... \\[0.5ex]
\hline
\multirow{2}{*}{$[n00]$}  &
    $A_1$  & $0^{(+)}$, 4 & $\bm 0^+$, $\bm 1^-$, $2^+$, ... & ...\\
   &$E_2$  & 1, 3 & $\bm 1^-$, $2^+$, ...                & $\bm 1^+$, ...\\[0.5ex]
   \hline
\multirow{2}{*}{$[nn0]$}  &
    $A_1$  & $0^{(+)}$, 2, 4 & $\bm 0^+$, $\bm 1^-$, $2^+_{[2]}$, ... & ...\\
   &$B_2,\:B_2$  & 1, 3 & $\bm 1^-$, $2^+$, ... & $\bm 1^+$, ...\\[0.5ex]
      \hline
\multirow{1}{*}{$[nnn]$}  &
    $A_1$  & $0^{(+)}$, 3    & $\bm 0^+$, $\bm 1^-$, $2^+$, ... & ...
\end{tabular}
\caption{The lowest continuum $J^P$ and helicity $\lambda$, and corresponding $D\pi$ (and similarly $D\eta$ and $D_s\bar{K}$) and $D^\ast\pi$ that subduce in each of the irreps and little groups used in this calculation. Overall momentum is denoted by $\vec{P}=\tfrac{2\pi}{L}(i,j,k)=[ijk]$. Bold $J^P$ denote a contribution that was used to obtain scattering information. $[N]$ indicates the number of subductions when more than one are present in an irrep and $\tilde\eta=P(-1)^J$. ``...'' denotes higher partial wave contributions which are not considered in this calculation. For each $J^P$ several $D^\star\pi$ combinations can appear, we only make use of $^{2S+1}\ell_J=\,\!^{3}\!S_1$ and ignore $P$-wave and higher. For $D\pi$ we consider up to $J^P=2^+$. Further details with more irreps and partial waves can be found in table 3 of Ref.~\cite{Wilson:2014cna} for $D\pi$ (which follows the same pattern as $K\pi$), and tables 1, 5, 6 and 7 of Ref.~\cite{Woss:2018irj} for $D^\star\pi$.}
\label{table:pw_irreps}
\end{center}
\end{table}

Correlation functions are computed using the distillation framework~\cite{Peardon:2009gh}, whereby the quark field is projected into a low dimensional space spanned by the lowest $N_{\text{vec}}$ eigenvectors of the gauge-covariant Laplacian ordered by eigenvalue. This projection suppresses high energy modes, enhancing the overlap of the operators onto the states we investigate. At the same time it provides an efficient way to compute correlators and reuse them for different operator constructions. For this analysis we use $N_{\text{vec}} = 256$ distillation vectors.\footnote{The same number of vectors was used in a recent study of $DK$ scattering in Ref.~\cite{Cheung:2020mql}; $N_\mathrm{vec}=384$ vectors were used in earlier studies of $\pi\pi$~\cite{Wilson:2015dqa,Briceno:2016mjc} and $\pi K$~\cite{Wilson:2019wfr} scattering on this lattice.}

In table~\ref{table:mesons} we summarise the stable hadron masses and the relevant thresholds for this calculation. These are obtained from the dispersion relation,
\begin{align}
(a_t E)^2 = (a_t m)^2 + {|\vec{d}|}^2\left(\frac{2\pi}{\xi \;L/a_s}\right)^2
\end{align}
where $m$ is the hadron mass, and $\vec{d}$ is a vector of integers. The anisotropy $\xi$ obtained from the pion dispersion relation is $\xi_\pi=3.453(6)$ and from the $D$ is $\xi_D=3.443(7)$. We use $\xi_\pi$ including its uncertainty to transform the moving-frame spectra to the rest-frame energies $E_\cm$. Both $\xi_\pi$ and $\xi_D$ are used subsequently to assess the uncertainties in the scattering amplitudes.

\begin{table}
\begin{center}
\begin{minipage}[c]{0.3\textwidth}
\begin{tabular}{c|c}
            & {$a_t m$} \\
\hline
$\pi$       & 0.03928(18)~\cite{Wilson:2015dqa}   \\
$K$         & 0.08344(7) ~\cite{Wilson:2015dqa}   \\
$\eta$      & 0.09299(56)~\cite{Wilson:2015dqa}   \\
$D$         & 0.30923(11)~\cite{Cheung:2016bym}   \\
$D_s$       & 0.32356(12)~\cite{Cheung:2020mql}   \\
$D^\ast$    & 0.33058(24)~\cite{Cheung:2020mql}   \\
\end{tabular} 
\end{minipage}  
\hspace{0.05\textwidth}
\begin{minipage}[c]{0.3\textwidth}
\begin{tabular}{c|c}       
                    & {$a_t E_\mathrm{threshold}$} \\
\hline
$D\pi$              & 0.34851(21) \\
$D\pi\pi$           & 0.38779(27)   \\
$D\eta$             & 0.40222(57)   \\
$D_s \bar{K}$       & 0.40700(14)   \\
$D^\ast \pi\pi$     & 0.40914(35)   \\
\end{tabular} 
\end{minipage}
\end{center}
\caption{Left: A summary of the stable hadron masses relevant for this calculation. Right: kinematic thresholds relevant for $I=1/2$ $D\pi$ scattering.}
\label{table:mesons}
\end{table}

\subsection{Determining the scattering amplitudes}
\label{sec:calc:amps}

The mapping between the quantised finite-volume spectrum obtained from lattice QCD, and the infinite volume scattering amplitudes is given by the L\"uscher quantisation condition~\cite{Luscher:1986pf,Luscher:1990ux,Luscher:1991cf}, and extensions thereof~\cite{Rummukainen:1995vs,Kim:2005gf,Christ:2005gi,Fu:2011xz,Leskovec:2012gb,Hansen:2012tf,Briceno:2012yi,Guo:2012hv,Briceno:2014oea}. We use the form,
\begin{equation}
\det \Bigl[ \bm{1} + i \bm{\rho}(s) \cdot \bm{t}(s) \cdot \bigl( \bm{1} + i\bm{\mathcal{M}}(s, L)\bigr) \Bigr] = 0,
\label{eq_det}
\end{equation}
where $s=E_\cm^2$, $\bm{\rho}$ is a diagonal matrix of phase space factors, $\rho(s)=2k(s)/\sqrt{s}$, $\bm{t}(s)$ is the infinite volume $t$-matrix which is diagonal in partial waves, and is related to the scattering $S$-matrix through $\bm{S}=\bm{1}+2i\sqrt{\bm{\rho}}\cdot \bm{t}\cdot\sqrt{\bm{\rho}}$. The determinant is over partial waves and open channels. The scattering momentum, $k(s)=\sqrt{\left(s-(m_D+m_\pi)^2\right)\left(s-(m_D-m_\pi)^2\right)/(4s)}$, and
${\bm{\mathcal{M}}}(s,L)$ is a matrix of known functions that encode the effect of the finite volume and mix partial waves. A more complete description including the subduction of this equation into lattice irreps, which is relevant for pseudoscalar-pseudoscalar scattering, with the matrix indices exposed, is given in Ref.~\cite{Wilson:2014cna}.

The solutions of Eq.~\ref{eq_det} for a given ${\bm t}(s)$ correspond to the finite-volume spectrum for that specific scattering amplitude ${\bm t}(s)$. To determine the scattering amplitude from a spectrum we choose to parameterise ${\bm t}(s)$ using several amplitudes that respect the unitarity of the $S$-matrix and are analytic except for cuts due to $k(s)$ and poles. The free parameters in ${\bm t}(s)$ can then be found by minimising a $\chi^2$ to best describe the spectrum obtained in the lattice calculation. 
The energies that solve Eq.~\ref{eq_det} can be identified by numerically root-finding the determinant, 
or the eigenvalues of the matrix inside the determinant~\cite{Woss:2020cmp}, as a function of $s=E_\mathsf{cm}^2$. 
These solutions are then used in a correlated $\chi^2$ fit as defined in Eq.~8 of Ref.~\cite{Wilson:2014cna}. 
Suitable parameterisations of $t$ explored here include $K$-matrices, the effective range expansion, a Breit-Wigner form, and 
unitarised chiral amplitudes.

We note that Eq.~\ref{eq_det} is valid only for two-hadron scattering processes. The choice to work at heavier-than-physical pion masses raises three-hadron (and higher) thresholds relative to two-hadron thresholds. To go rigorously beyond three-hadron 
thresholds such as $D\pi\pi$ which appears at relatively low energies for the pion masses used here, 
an extension to the theoretical framework is required~\cite{Hansen:2019nir}.  The three-body quantisation condition has 
been applied recently to simple systems of three identical particles~\cite{Horz:2019rrn,Blanton:2019vdk,Mai:2019fba,Culver:2019vvu,Fischer:2020jzp,Hansen:2020otl} while a very recent study has reported the generalisation of the quantisation condition to three non-identical particles~\cite{Blanton:2020gmf}. While this is promising progress, a full consideration of the energy region above three-particle thresholds is beyond the scope of the current work.  

\section{Finite Volume Spectra}
\label{sec:spectra}

\begin{figure}[tb]
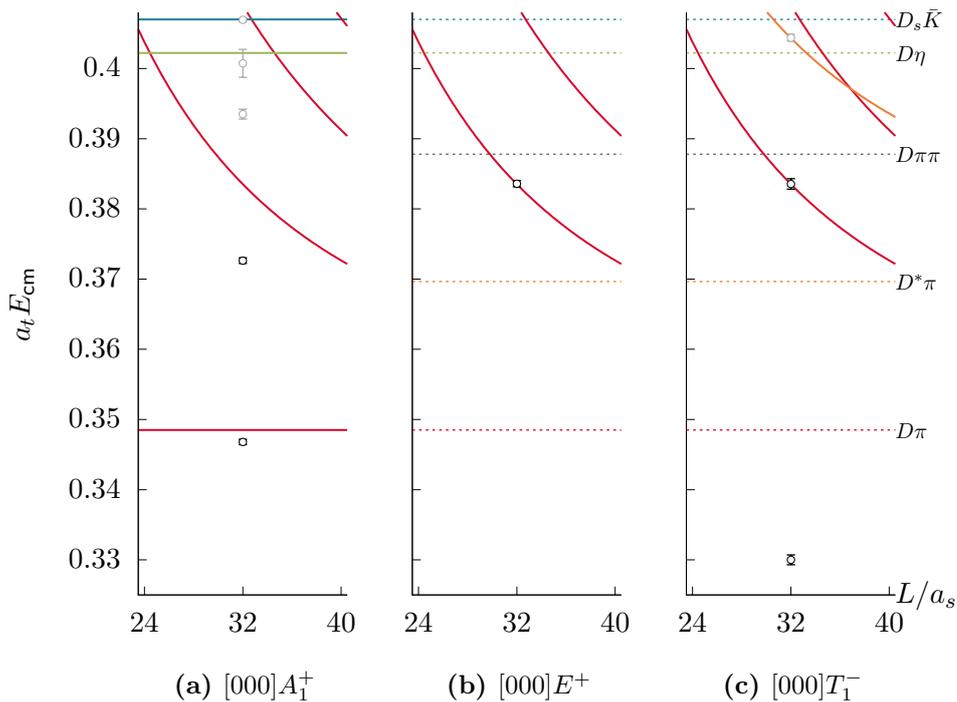

	\centering
	\begin{subfigure}[l]{0.23\textwidth}
		\hspace*{-1cm}
		\graphicspath{{plots/irreps/}}
		\input{plots/irreps/000_A1p_pub.tex}
		\subcaption{$[000] A_1^+$}
	\end{subfigure}
	\begin{subfigure}[l]{0.23\textwidth}
		\hspace*{-1cm}
	\graphicspath{{plots/irreps/}}
	\input{plots/irreps/000_Ep_pub.tex}
	\subcaption{$[000] E^+$}
	\end{subfigure}
	\begin{subfigure}[l]{0.23\textwidth}
		\hspace*{-1cm}
		\graphicspath{{plots/irreps/}}
		\input{plots/irreps/000_T1m_pub.tex}
		\subcaption{$[000] T_1^-$}
	\end{subfigure}
	\caption{Finite-volume spectra obtained in the at-rest $[000] A_1^+$, $[000] E^+$ and $[000] T_1^-$ irreps. Dotted lines correspond to channel thresholds. Solid lines indicate non-interacting energy levels corresponding to operators included in the simulation. Points with error bars represent the energy levels obtained from the variational analysis. Black points will be included in the subsequent scattering analysis while grey points will be excluded.}
\label{fig:spec1}
\end{figure}

\begin{figure}[tb]
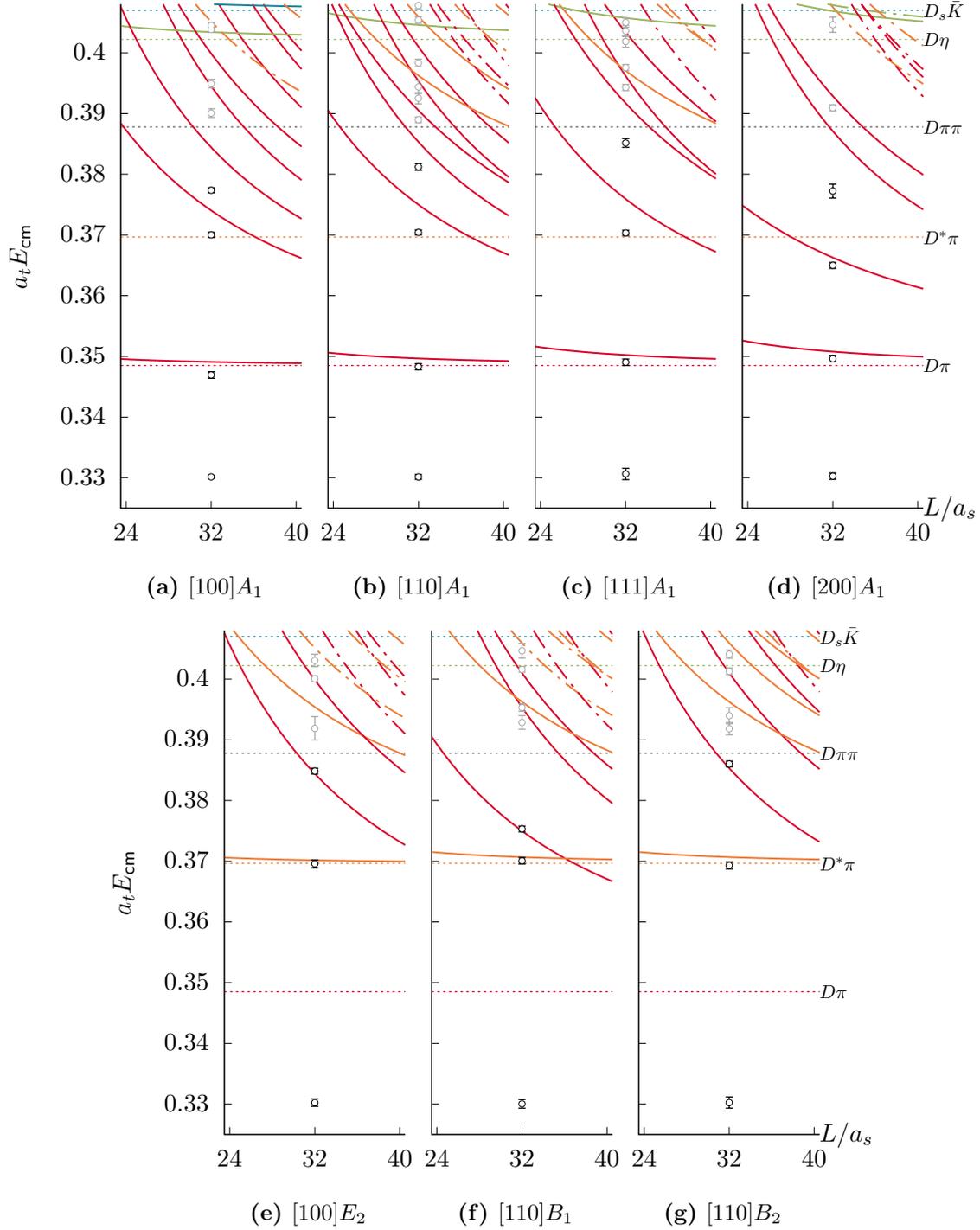

	\centering
	\begin{subfigure}[l]{0.20\textwidth}
		\hspace*{-1.1cm}
		\graphicspath{{plots/irreps/}}
		\input{plots/irreps/100_A1_pub.tex}
		\subcaption{$[100] A_1$}
	\end{subfigure}
	\begin{subfigure}[l]{0.20\textwidth}
		\hspace*{-1.1cm}
		\graphicspath{{plots/irreps/}}
		\input{plots/irreps/110_A1_pub.tex}
		\subcaption{$[110] A_1$}
	\end{subfigure}
	\begin{subfigure}[l]{0.20\textwidth}
		\hspace*{-1.1cm}
		\graphicspath{{plots/irreps/}}
		\input{plots/irreps/111_A1_pub.tex}
		\subcaption{$[111] A_1$}
	\end{subfigure}
	\begin{subfigure}[l]{0.20\textwidth}
		\hspace*{-1.1cm}
		\graphicspath{{plots/irreps/}}
		\input{plots/irreps/200_A1_pub.tex}
		\subcaption{$[200] A_1$}
	\end{subfigure}
	\begin{subfigure}[l]{0.20\textwidth}
		\hspace*{-1.1cm}
		\graphicspath{{plots/irreps/}}
		\input{plots/irreps/100_E2_pub.tex}
		\subcaption{$[100] E_2$}
	\end{subfigure}
	\begin{subfigure}[l]{0.20\textwidth}
		\hspace*{-1.1cm}
		\graphicspath{{plots/irreps/}}
		\input{plots/irreps/110_B1_pub.tex}
		\subcaption{$[110] B_1$}
	\end{subfigure}
	\begin{subfigure}[l]{0.20\textwidth}
		\hspace*{-1.1cm}
		\graphicspath{{plots/irreps/}}
		\input{plots/irreps/110_B2_pub.tex}
		\subcaption{$[110] B_2$}
	\end{subfigure}
	\caption{As in Fig.~\ref{fig:spec1}, but for the moving-frame $[100] A_1$, $[110] A_1$, $[111] A_1$, $[200] A_1$ irreps (top) and $[100] E_2$, $[110] B_1$ and $[110] B_2$ irreps (bottom). The dash-dotted curves indicate a non-interacting level for which no corresponding operator was included in the basis.}
\label{fig:spec2}
\end{figure}

The finite volume spectra computed in this study are presented in Figs.~\ref{fig:spec1} and \ref{fig:spec2}. As explained above, these spectra are grouped according to the irreps of the cubic group and little groups, labelled by $[\vec{d}] \Lambda^{(P)}$, where $\vec{d}$ indicates the direction of the overall momentum such that $\vec{P} = 2\pi \vec{d}/L$. Parity $P$ is only a valid quantum number at zero overall momentum. Energy levels used to constrain the scattering amplitudes are shown in black, while other energies extracted but not used are shown in grey. The cutoff for energies used in the scattering analyses is $D\pi\pi$ threshold. 

We compute correlation functions on a single volume in this analysis, but we nonetheless indicate the volume dependence of the non-interacting levels in these plots. Only certain continuum angular momenta subduce into a given irrep in the cubic volume as indicated in table~\ref{table:pw_irreps}, and higher partial waves are suppressed by a factor of $k^{2\ell+1}$ near threshold. In Fig.~\ref{fig:spec1}, the at-rest irreps are shown. The $D\pi$ $S$-wave ($\ell=0$) only subduces into $[000]A_{1}^{+}$ in this figure. The $D\pi$ $P$-wave ($\ell=1$) subduces into $[000]T_{1}^{-}$, while the lowest contribution for $[000]E^{+}$ is the $D \pi$ $D$-wave ($\ell=2$), as summarised in table~\ref{table:pw_irreps}. $[000]E^{+}$ will not be used in any of the fits as discussed in section~\ref{sec:scattering}. The irreps shown in Fig.~\ref{fig:spec2} have non-zero total momentum. The $D \pi$ $S$-wave subduces into the top four of these ($[100] A_1$, $[110] A_1$, $[111] A_1$ and $[200] A_1$). 
The lower three irreps, $[100]E_2$, $[110]B_1$ and $[110]B_2$, predominantly contain the $D\pi$ $P$-wave close to threshold and the $D^* \pi$ $S$-wave at slightly higher energies and will be considered together with $[000]T_{1}^{-}$, but separately from the $D\pi$ $S$-wave.

All irreps that have an $\ell=1$ contribution have a level far below threshold which may be associated with a deeply bound $D^\star$ vector state that is stable at this heavier-than-physical light quark mass. In all irreps with an $\ell=0$ contribution we observe a level around $D\pi$ threshold that is shifted downward with respect to the nearby non-interacting level. We also observe the appearance of what may be an extra level around $a_t E_\cm = 0.37$ and an upward shift of higher levels with respect to their non-interacting energies. This is an indicator for non-trivial $S$-wave interactions. In comparison, irreps having $\ell=1$ as the lowest partial wave contribution yield levels which are only marginally shifted away from the nearby non-interacting energies.

\begin{figure}[tb]
\centering
\includegraphics[width=0.9\textwidth]{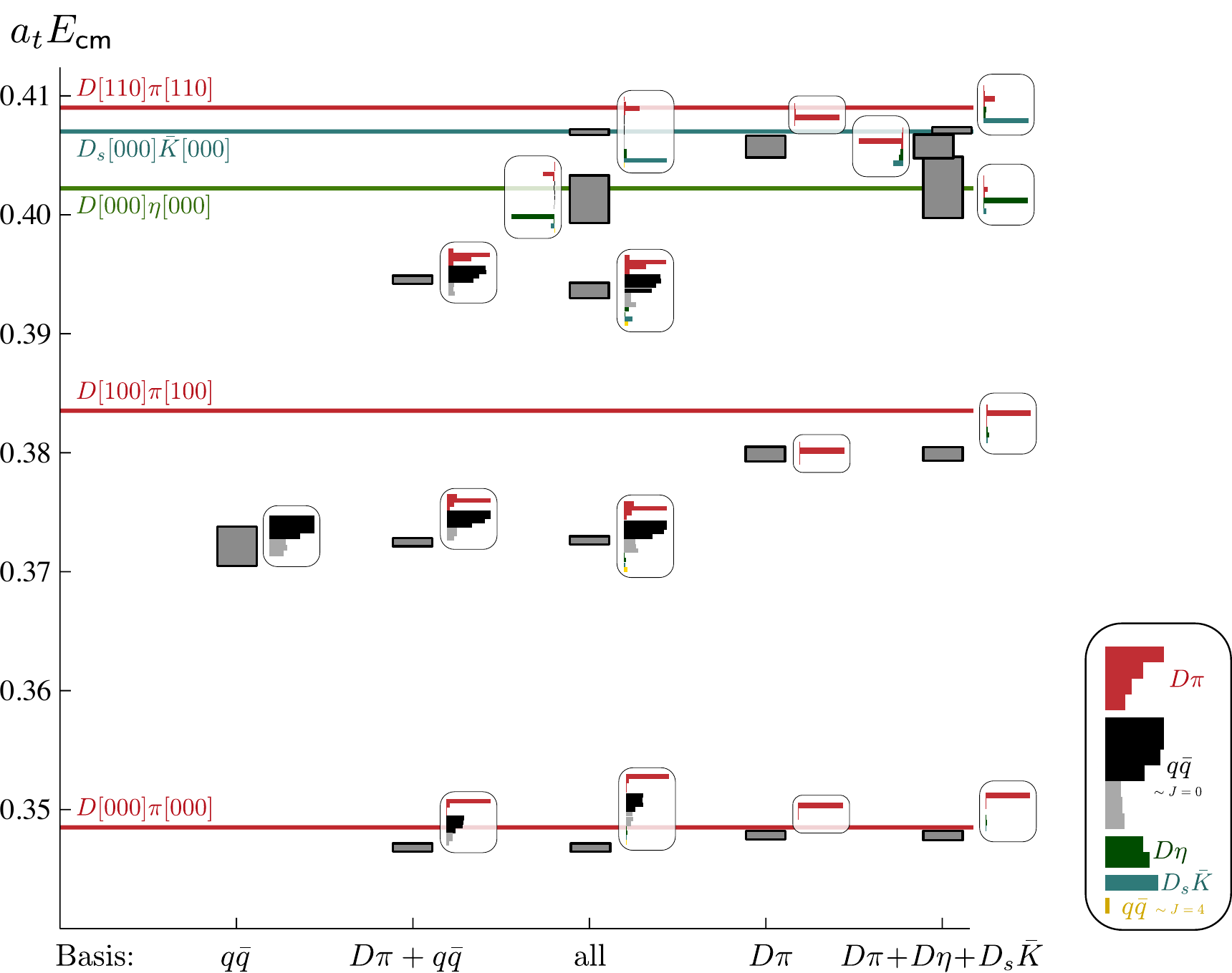}
\caption{The spectra obtained in $[000]A_1^+$ when varying the operator basis. The operators included are marked below each column.  The grey blocks show the $1\sigma$ uncertainties of the energies obtained from variational method. The lowest two energies from the column marked ``all'' correspond to the levels used in the scattering analyses. The histograms plotted next to each energy are the magnitudes of the operator overlaps $\left<0|\mathcal{O}|\mathfrak{n}\right>$, normalised to their maximum contribution seen in any state, from each variational analysis. The solid lines indicate the non-interacting energies.}
\label{fig:basis}
\end{figure}

Figure \ref{fig:basis} shows the spectrum obtained for $[000]A_{1}^{+}$ and how this can vary when different types of operator are removed. It is clear from the figure that neither the $D \pi$-like nor the $q\bar{q}$-like operators alone are enough to reliably compute the spectrum in this system. 
Using the $q \bar{q}$ operators alone produces a single level at a similar energy to the second level found when using a more complete basis, however the lowest level close to $D[000]\pi[000]$ is absent. Using only $D\pi$-like operators does not produce a single level consistent with the spectrum found when using the more complete basis. Adding the $D \eta$-like and the $D_s \bar{K}$-like operators leaves the spectrum unchanged at lower energies and new levels arise close to the non-interacting levels for $D[000]\eta[000]$ and $D_s[000]\bar{K}[000]$. The lowest two levels used in the scattering analyses are robust against any small changes in the operator basis, such as removing higher lying operators and many individual $q\bar{q}$-like operators.

While qualitative indications of the interactions present can be obtained from looking at the finite volume spectra, in order to gain a rigorous understanding the energies obtained must be related to infinite volume scattering amplitudes, as we do in the next section.

\section{Scattering analyses}\label{sec:scattering}
In this section, the spectra presented above are used to constrain the infinite volume scattering amplitudes through the L\"uscher determinant condition given by Eq.~\ref{eq_det}. The goal is to extract the $D\pi$ $S$-wave. However due to the mixing of angular momenta introduced by the cubic volume it is necessary to consider other partial waves subducing into any of the irreps used in the analysis. As mentioned before, partial waves grow at threshold like $k^{2\ell+1}$ which means that higher $\ell$ are suppressed close to threshold.\footnote{In the absence of a nearby resonance or bound-state pole.} In this calculation we only find effects of any significance from $\ell=0$ and $\ell=1$.

At zero overall momentum $[000]A_1^+$ is the only irrep with an $S$-wave contribution, as is shown in table~\ref{table:pw_irreps}. The next higher partial wave that subduces in this irrep has $\ell=4$ and can be neglected. However there are not sufficiently many energy levels to constrain the amplitude from this irrep alone. In the $A_1$ irreps at non-zero momentum $D\pi$ can contribute with $J^P=0^+,1^-,2^+, ...$ and thus it is necessary to consider these partial waves simultaneously. The lowest $D^\star \pi$ non-interacting level in these irreps is $D^\star_{[100]} \pi_{[100]}$ in $[110]A_1$ which lies well above the $D\pi\pi$ threshold, and the lowest contributing partial wave is $D^\star\pi$ in $P$-wave. Hence $D\pi$ is the only relevant channel in these irreps within the energy region we consider.
In the moving frame $[100]E_2$ and $[110]B_{1,2}$ irreps with nonzero total momentum, $J^P=1^-$ $D\pi$ scattering is the leading partial wave. These irreps also include a contribution from $D^\star \pi$ in $J^P=1^+$ and the lowest non-interacting level above $D\pi$ threshold is $D^\star_{[100]} \pi_{[000]}$. We therefore need to include $D^\star \pi$ in the $t$-matrix when making use of energy levels in these irreps.

On the basis of these observations we initially perform two separate amplitude determinations: One using $[000]T_1^-$, $[100]E_2$, and $[110]B_{1,2}$ to constrain the $D\pi$ $P$-wave and to assess the $D^\star \pi$ $S$-wave contribution (section \ref{sec:scattering:P}). The second determination uses $[000]A_1^+$ and the moving frame $A_1$ irreps to constrain the $D\pi$ $S$-wave and $P$-wave simultaneously (section \ref{sec:scattering:SP}). Treating energies separately simplifies the $D\pi$ $S$-wave analysis, while retaining constraints on the $D\pi$ $P$-wave. 

In this analysis, we only consider the region below the three-body $D\pi\pi$ threshold at $a_t E_\cm \approx 0.388$, which is also well below the $D \eta$ threshold. No levels in this energy region show sensitivity to the inclusion of $D\eta$ operators. Effects of higher partial waves with $\ell \geq 2$ were also investigated and found to be negligible in all fits we perform. $[000]E^+$ is an irrep where $\ell=2$ $D\pi$ is the lowest partial wave. In this irrep, the lowest level is found to be consistent with the lowest non-interacting level, $D[100]\pi[100]$, at $a_tE_\cm= 0.38333\pm 0.00049$, and corresponds to a $D$-wave scattering phase $\delta_2 = (0.49 \pm 1.29)^\circ$. Since the phase must be zero at threshold it is reasonable to conclude that the $D\pi$ $D$-wave is negligibly small throughout the energy region used for scattering analyses.

\subsection{Parameterising the $t$-matrix }
We now introduce the $t$-matrix parameterisations used in this analysis. In elastic pseudoscalar-pseudoscalar scattering there is no coupling between partial waves in an infinite volume, the $t$-matrix is therefore diagonal in partial waves. In this case, a single partial wave amplitude can be described by the scattering phase shift $\delta_\ell(E_{\cm})$ related to the $t$-matrix through $t^{(\ell)} = \frac{1}{\rho} e^{i \delta_\ell} \sin \delta_\ell$.\footnote{This is also true of the $J^P=1^+$ $D^\star\pi$ amplitude we consider, assuming only $^{2S+1}\ell_J=\:\!^3S_1$ is relevant close to threshold, neglecting $^{2S+1}\ell_J=\:\!^3D_1$.}

Ultimately the results obtained should not be dependent on the intermediate parameterisation used to described the $t$-matrix, and relying on only a single expression for $t^{(\ell)}(s)$ may introduce bias. We thus parameterise $t^{(\ell)}(s)$ in a variety of ways. A flexible parameterisation respecting unitarity of the $S$-matrix is a $K$-matrix,\footnote{Although referred to as matrices in general, in this calculation these are scalar equations for each partial wave amplitude.} which for elastic scattering is given by
\begin{equation}
(t^{(\ell)})^{-1}(s) = \frac{1}{(2k)^\ell} K^{-1}(s) \frac{1}{(2k)^\ell} + I(s) \, ,
\end{equation}
for a partial wave $\ell$, and $K(s)$ that is real for real values of $s$. The factors $(2k)^{-\ell}$ ensure the expected threshold behaviour. Unitarity of the $S$-matrix is guaranteed if $\mathrm{Im}\:I(s)=-\rho(s)$ above threshold. This places no constraint on Re$\:I(s)$. One choice is to set Re$\:I(s)$ to zero above threshold, giving $I(s) = -i \rho(s)$. Another option is to use the Chew-Mandelstam prescription \cite{Chew:1960iv}, which uses the known Im$\:I(s)$ to generate a non-zero Re$\:I(s)$ through a dispersion relation, whose explicit form is given in appendix B of Ref.~\cite{Wilson:2014cna}. To make the dispersion relation integral converge a subtraction at an arbitrary value of $s$ is needed.

A general expression for the amplitudes we use in sections~\ref{sec:scattering:P} and \ref{sec:scattering:SP} is
\begin{equation}
K(s) =  \frac{\left( g^{(0)} + g^{(1)}s \right)^2}{m^2 - s} + \gamma^{(0)} + \gamma^{(1)} s \, ,
\label{eq_K_general}
\end{equation}
where $g^{(n)}$, $\gamma^{(n)}$ and $m$ are real free parameters that are obtained by the minimisation procedure described in section~\ref{sec:calc:amps}. Various parameters may be fixed to zero for different applications. When a $K$-matrix pole parameter is present, as in Eq.~\ref{eq_K_general}, when using a Chew-Mandelstam phase space we subtract at the pole parameter, $s=m^2$.

In section~\ref{sec:scattering:var} we make use of variations of Eq.~\ref{eq_K_general} and, additionally, of a ratio of polynomials,
\begin{align}
K^{-1}(s)=\frac{\sum_{n=0}^N c_n s^n}{1+\sum_{m=1}^M d_m s^m}\,.
\label{eq_K_ratio}
\end{align}
where $c_n$ and $d_m$ are real free parameters. Several of the low-order truncations of Eq.~\ref{eq_K_ratio} are algebraically identical to Eq.~\ref{eq_K_general}, however parameter correlations can differ significantly. One choice that often reduces correlations is the replacement $s\to\hat{s} \equiv (s-s_\mathrm{thr.})/s_\mathrm{thr.}$. When using Eq.~\ref{eq_K_ratio} with a Chew-Mandelstam phase space, we choose to subtract at threshold, $s=s_{\mathrm{thr.}}=(m_\pi+m_D)^2$.

In the case of single-channel elastic scattering, a common choice of amplitude is an effective range expansion, given by
\begin{equation}
	k^{2\ell+1} \cot \delta_\ell = \frac{1}{a_\ell} + \frac 1 2 r_\ell k^2 + P_2 k^4 + \mathcal{O}(k^6)\,,
	\label{eq_ER}
\end{equation}
where $a_\ell$ and $r_\ell$ are the scattering length and effective range respectively, for partial wave $\ell$. 

Another common parameterisation, that is appropriate for a single isolated resonance, is the relativistic Breit-Wigner parameterisation
\begin{equation}
t^{(\ell)}(s) = \frac{1}{\rho(s)} \frac{\sqrt{s} \Gamma_\ell(s)}{m_R^2 - s - i \sqrt{s} \Gamma_\ell(s)} \,,
\label{eq_BW}
\end{equation}
where the width is given by $\displaystyle \Gamma_\ell(s) = \frac{g_R^2}{6 \pi}\frac{k^{2\ell+1}}{s \, m_R^{2(\ell-1)}}$, which ensures the correct near-threshold behaviour, and $m_R$, $g_R$ are free parameters.

We also consider unitarised chiral amplitudes that have been applied several times to $D\pi$ scattering~\cite{Hofmann:2003je,Guo:2008gp,Guo:2009ct,Albaladejo:2016lbb,Guo:2018kno,Guo:2018tjx}. Chiral perturbation theory is an effective field theory (EFT) approach, derived from expanding a Lagrangian of meson degrees of freedom about the chiral ($m_u, m_d, m_s\to0$) and small-momentum limits. The number of terms grows order-by-order, and these come with unknown Wilson coefficients not specified by the EFT that are estimated either from experimental data or from a first-principles approach such as lattice QCD. In many cases of interest, when extrapolating these amplitudes away from threshold, they grow larger than permitted by unitarity of the $S$-matrix, necessitating unitarisation. This also enables resonance poles to be generated, which can otherwise be difficult to achieve through an expansion in momentum. Next-to-leading order is required for meson loops to appear, however, we choose to fix the next-to-leading order Wilson coefficients to zero to reduce the number of free parameters. The amplitude used is similar to the $K$-matrices described above, and is shown to be algebraically identical to Eq.~\ref{eq_K_ratio}, with $M=N=2$ and specific coefficients $c_n$ and $d_m$ that depend on $m_D$ and $m_\pi$, in Appendix~\ref{app:sec:chipt}. In addition to the analyticity and unitarity (in $s$) shared with all the $K$-matrix amplitudes used, this amplitude also has the assumption of chiral symmetry, that restricts the behaviour at threshold.

We use a simple form that can be written as
\begin{align}
K^{-1}(s)&=\left(-\frac{1}{16\pi}\mathcal{V}_{J=0}\right)^{-1}+\frac{\alpha(\mu)}{\pi} + \frac{2}{\pi}\left(\frac{m_D}{m_\pi+m_D}\log\frac{m_D}{m_\pi} + \log\frac{m_\pi}{\mu}\right)\;.
\label{eq_K_chipt}
\end{align}
with a threshold-subtracted Chew-Mandelstam phase-space $I(s)$, and
\begin{align}
\mathcal{V}_{J=0} &= -\frac{1}{4sF^2}\Bigl(3s^2-2s(m_D^2+m_\pi^2)-(m_D^2-m_\pi^2)^2\Bigr)
\end{align}
corresponds to the $S$-wave projected leading-order elastic $D \pi$ scattering amplitude. $F$ and $\alpha(\mu)$ are treated as free parameters, and the renormalisation scale $\mu$ is fixed to $a_t \mu = 0.1645$ corresponding to $\mu\approx 1000$~MeV in physical units.

\subsection{$D\pi$ with $J^P=1^-$ and $D^\star\pi$ with $J^P=1^+$}
\label{sec:scattering:P}

We begin with a fit of the $[000]T_1^-$, $[100]E_2$, $[110]B_1$ and $[110]B_2$ spectrum. These irreps contain $D\pi$ in $P$-wave but do not have a $D\pi$ $S$-wave contribution. There is a level far below threshold in $T_1^-$ that signals a $J^P=1^-$ $D^\star$ bound state. The moving frame irreps contain a contribution from $D^\ast \pi$ in $S$-wave ($J^P=1^+$). To parameterise these two partial waves, we use Eq.~\ref{eq_K_general}, once with a pole term in $J^P=1^-$ $D\pi$, and again with a constant in $J^P=1^+$ $D^\ast \pi$, and obtain
\begin{equation*}
\begin{aligned}[t]
\begin{matrix}
\gamma ^{(0)\:D^* \pi} &= &(1.35  \pm 0.83 \pm 0.45) \\
g_1^{D \pi} &= &(0.72  \pm 0.31 \pm 0.13)  \\
m_1 &= &(0.33028  \pm 0.00052 \pm 0.00005) \cdot a_t^{-1}
\end{matrix}
\end{aligned}
\qquad
\begin{aligned}[t]
\begin{bmatrix}
& & 1.00 & -0.72 & -0.41 \\
& & & 1.00 & 0.34 \\
& & & & 1.00 
\end{bmatrix}
\end{aligned}
\end{equation*}
\begin{equation}
	\chi^2/N_{\text{dof}} = \tfrac{8.59}{11 - 3} = 1.07\;.
	\label{eq_PmPp}
\end{equation}
The first uncertainty is obtained from the $\chi^2$ minimum, as is the matrix on the right that shows the parameter correlation.  The second uncertainty indicated is obtained by additional $\chi^2$ minimisations, after varying in turn the $\pi$, $D$ and $D^\star$ masses and anisotropy to the maximum and minimum values within their $1\sigma$ uncertainties, and taking the maximum deviation.\footnote{For a given parameter $x$ with central value $\bar{x}$, $x_i$ values are obtained for each mass and each anisotropy variation $i$, and uncertainties $\sigma_{x_i}$. Then the second uncertainty quoted is $\mathrm{max}_{i}\left( |\bar{x}\pm\sigma_{\bar{x}}| - |x_i\pm\sigma_{x_i}|\right)$ where the two $\pm$ are changed simultaneously. In this case the variations considered are $\{m_D\to m_D\pm\sigma_{m_D},\; m_{D^\star}\to m_{D^\star}\pm\sigma_{m_{D^\star}},\;m_\pi\to m_\pi\pm\sigma_{m_\pi},\; \xi\to\xi_\pi+\sigma_{\xi_\pi},\; \xi\to\xi_{D}-\sigma_{\xi_D}\}$, where the two anisotropy variations are the largest possible deviations from the mean for $\xi_\pi$. This procedure is repeated for the other minima highlighted below.}

\begin{figure}[tb]
	\centering

	\graphicspath{{plots/}}
	\input{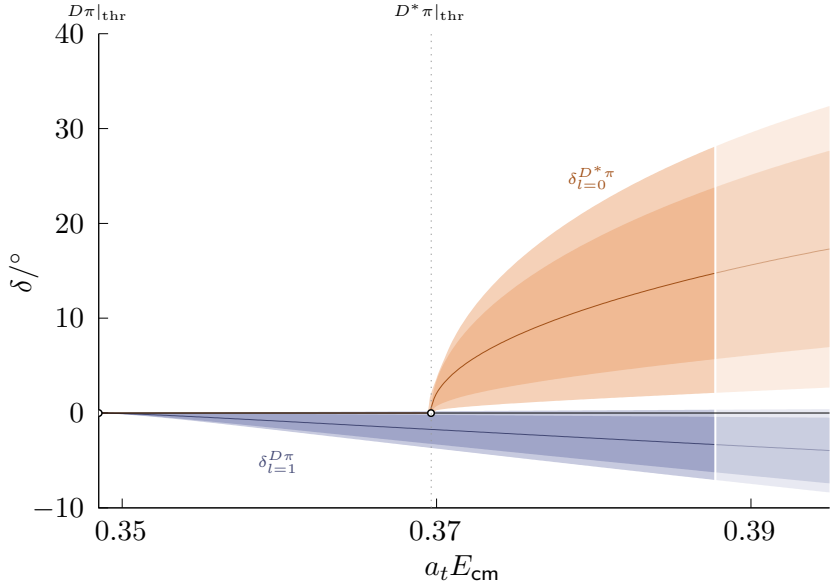}

	\caption{The phase shift for the $D\pi$ $P$-wave (blue) and $D^\ast \pi$ $S$-wave (orange) amplitudes. The inner band corresponds to the statistical uncertainties from the $\chi^2$-minimum in Eq.~\ref{eq_PmPp}. The outer band shows the maximum possible deviation when varying the scattering particle masses and anisotropy within their uncertainties. The faded region begins at $D\pi\pi$ threshold indicating the highest energy considered in this calculation.}
	\label{P_phase_shift}
\end{figure}

The phase shifts determined from this amplitude are shown in Fig.~\ref{P_phase_shift}. The inner bands correspond to the parameters in Eq.~\ref{eq_PmPp} using the first uncertainties and the correlations. The outer bands show the largest deviation determined by varying the mass and anisotropy values. The elastic $D \pi$ $P$-wave is small and in the next section we will also find a similar small $P$-wave phase shift in the combined fit of $D\pi$ $J^P=0^+$ and $J^P=1^-$. The $D^\star\pi$ contribution rises at threshold, perhaps indicating the tail of a higher $D_1$ resonance. Some evidence for this can be read off from Fig.~\ref{fig:spec2} where ``extra'' levels appear around $a_tE_\cm\approx0.39$. Additionally we include a simultaneous fit of $J^P=0^+,1^-$ $D\pi$ and $J^P=1^+$ $D^\star\pi$ in appendix~\ref{app:sec:fullfit}, where the resulting $J=1$ amplitudes are very similar to those in Eq.~\ref{eq_PmPp} and Fig.~\ref{P_phase_shift}.

\subsection{$D\pi$ with $J^P=0^+$ and $J^P=1^-$}
\label{sec:scattering:SP}

We now determine the $S$ and $P$-wave amplitudes simultaneously using 20 energy levels below $E = m_D + 2m_{\pi}$ from the $[000]A_1^+$, $[000]T_1^-$, and the four moving-frame $A_1$ irreps. We begin by fitting a ``reference'' amplitude, which consists of an $S$-wave $K$-matrix with a pole term and a constant $\gamma$, and a $P$-wave with just a pole term, each as defined in Eq.~\ref{eq_K_general}. Both use a Chew-Mandelstam phase-space subtracted at $s=m^2$ in each partial wave. After fitting these five free parameters, we find
\begin{align}
&
\begin{aligned}[t]
\begin{matrix}
m &= &(0.401  \pm 0.010 \pm 0.007) \cdot a_t^{-1} \\
g &= &(0.419 \pm 0.083 \pm 0.066) \cdot a_t^{-1} \\
\gamma^{(0)} &= &(-2.0  \pm 1.3 \pm 0.9) \\
m_1 &= &(0.33018 \pm 0.00016 \pm 0.00002) \cdot a_t^{-1} \\
g_1 &= &(0.63  \pm 0.51 \pm 0.30)  \\
\end{matrix}
\end{aligned}
\qquad
\begin{aligned}[t]
\begin{bmatrix}
1.00 & 0.93 & -0.62 & 0.23 & -0.10 \\
& 1.00 & -0.85 & 0.17 & 0.05 \\
& & 1.00 & -0.08 & -0.30 \\
& & & 1.00 & -0.10 \\
& & & & 1.00 
\end{bmatrix}
\end{aligned}\nonumber
\\
&\qquad\qquad\qquad\qquad\qquad\qquad\chi^2/N_{\text{dof}} = \tfrac{13.49}{20 - 5} =  0.90\,,
\label{eq_ref_amp}
\end{align}
where the parameters with a subscript ``1'' describe the $P$-wave, and those without describe the $S$-wave. The meaning of the uncertainties are as described below Eq.~\ref{eq_PmPp}.

\begin{figure}[tb]
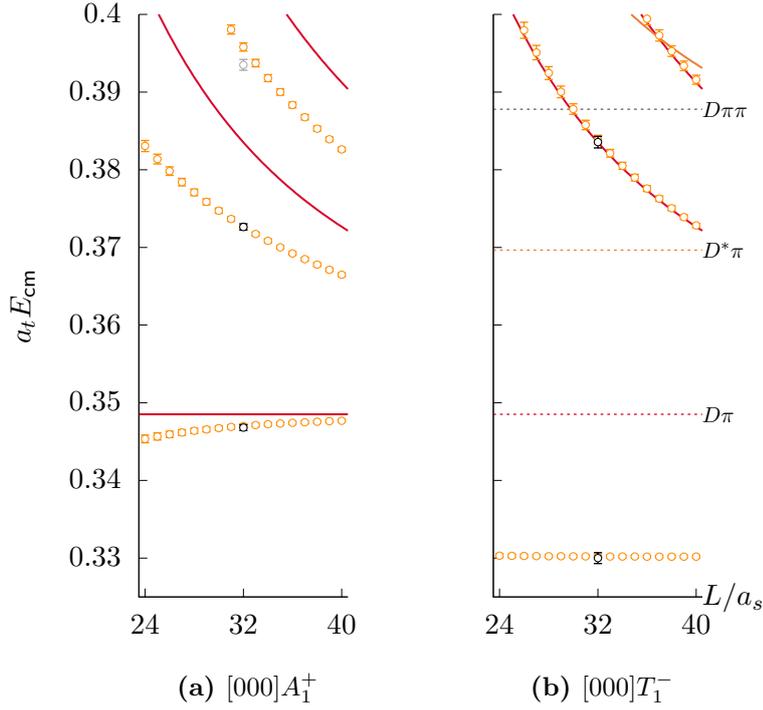

	\centering
	\begin{subfigure}[l]{0.30\textwidth}
		\hspace*{-0.5cm}
		\graphicspath{{plots/irreps/}}
		\input{plots/irreps/000_A1p_fvs_pub2.tex}
		\subcaption{$[000] A_1^+$}
	\end{subfigure}
	\begin{subfigure}[l]{0.30\textwidth}
		\hspace*{-0.5cm}
		\graphicspath{{plots/irreps/}}
		\input{plots/irreps/000_T1m_fvs_pub2.tex}
		\subcaption{$[000] T_1^-$}
	\end{subfigure}
	\caption{Finite-volume spectra obtained in the at-rest $A_1^+$ and $T_1^-$ irreps, as in Fig.~\ref{fig:spec1}, plotted with the solutions of the L\"uscher determinant condition using the reference parameterisation with the parameters resulting from the $\chi^2$-minimisation (orange points).}
\label{fvs_at_rest}
\end{figure}

\begin{figure}[tb]
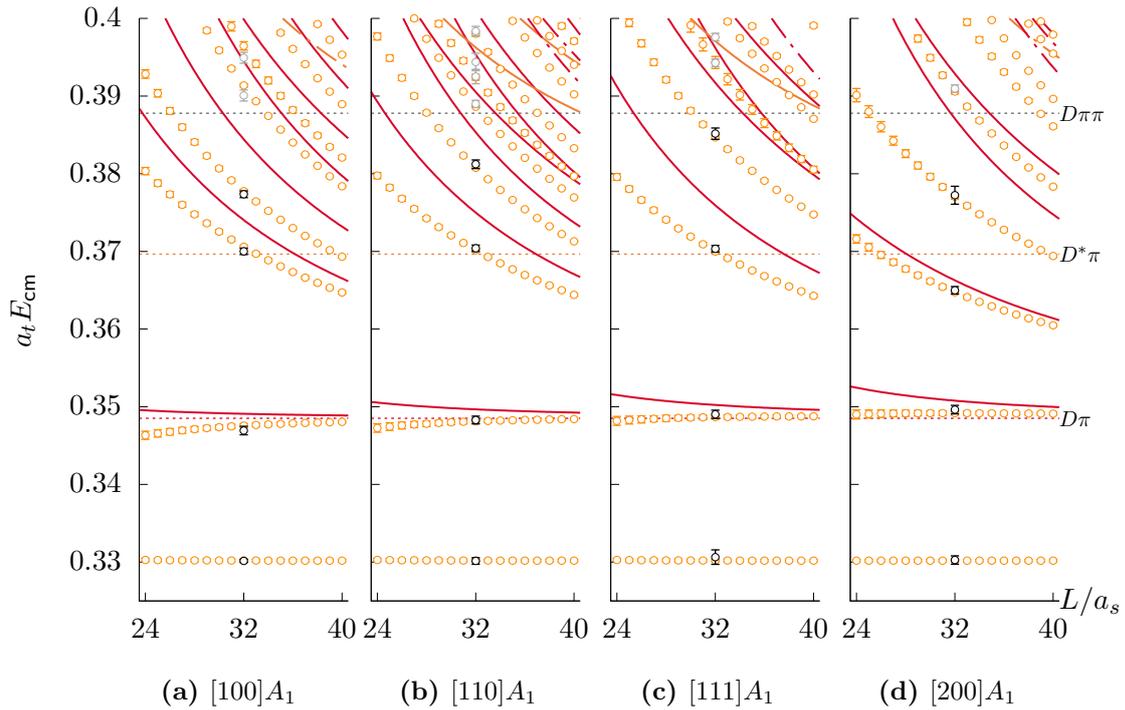

\centering
\begin{subfigure}[l]{0.20\textwidth}
	\hspace*{-1cm}
\graphicspath{{plots/irreps/}}
\input{plots/irreps/100_A1_fvs_pub2.tex}
\subcaption{$[100] A_1$}
\end{subfigure}
\begin{subfigure}[l]{0.20\textwidth}
	\hspace*{-1.1cm}
\graphicspath{{plots/irreps/}}
\input{plots/irreps/110_A1_fvs_pub2.tex}
\subcaption{$[110] A_1$}
\end{subfigure}
\begin{subfigure}[l]{0.20\textwidth}
	\hspace*{-1.1cm}
\graphicspath{{plots/irreps/}}
\input{plots/irreps/111_A1_fvs_pub2.tex}
\subcaption{$[111] A_1$}
\end{subfigure}
\begin{subfigure}[l]{0.20\textwidth}
	\hspace*{-1.1cm}
\graphicspath{{plots/irreps/}}
\input{plots/irreps/200_A1_fvs_pub2.tex}
\subcaption{$[200] A_1$}
\end{subfigure}
\caption{As Fig.~\ref{fvs_at_rest}, but for the moving frame $A_1$ irreps.}
\label{fvs_moving}
\end{figure}

Figures \ref{fvs_at_rest} and \ref{fvs_moving} show comparisons of the spectra obtained from the lattice calculation and the spectra obtained from the solutions of the L\"uscher determinant condition using the reference parameterisation Eq.~\ref{eq_ref_amp}. There is good agreement below $a_t E_\cm \approx 0.39$.

\begin{figure}[tb]
	\centering
		\graphicspath{{plots/}}
		\input{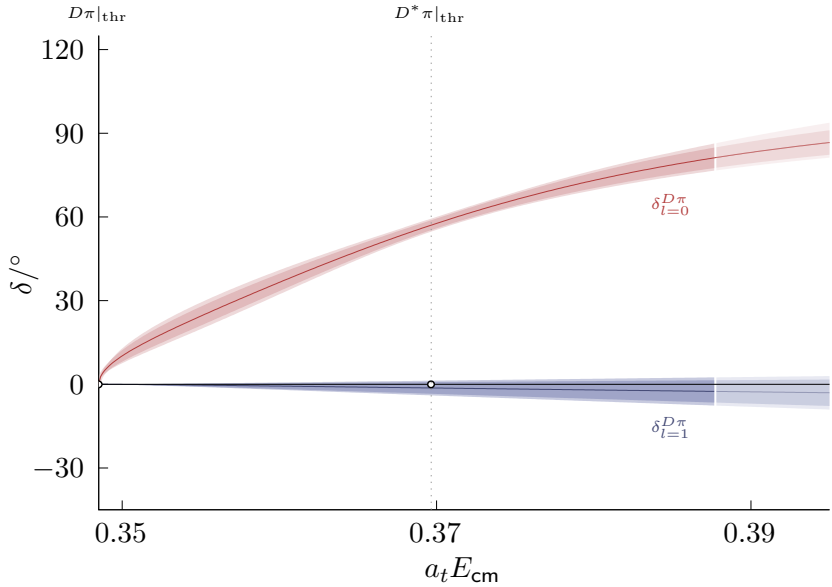}
	\caption{Phase shift of the $S$-wave (red) and $P$-wave (blue) $D\pi$ amplitudes.
		The inner line corresponds to the reference parameterisation. The inner dark error band represents the statistical error from the $\chi^2$-minimisation while the outer light error band additionally includes uncertainties from varying the input hadron masses and anisotropy within $1 \sigma$.}
	\label{SP_phase_shift}
\end{figure}

In Fig.~\ref{SP_phase_shift}, we show the phase shift of both the $D\pi$ $S$-wave and $P$-wave. The $S$-wave amplitude turns on rapidly at threshold and rises monotonically towards the edge of the elastic region, which in the finite volume produces statistically significant energy shifts and perhaps an additional level, suggestive of a resonance. We plot this $S$-wave amplitude again as $k\cot\delta_0$ in Fig.~\ref{fig_860_ere} and the $P$-wave as $k^3\cot\delta_1$ in Fig.~\ref{fig_P_ere} in appendix~\ref{app:kcot}. We defer the discussion of the poles and thus the resonance content of the $t$-matrix until after we have considered varying the form of the parameterisation.

\subsection{Parameterisation variations} 
\label{sec:scattering:var}

We now consider a range of parameterisations to explore the sensitivity to any particular choice. We perform minimisations to two different selections of energies. Motivated by the lack of volume dependence of the deeply bound levels due to the vector $D^\star$ state, and small $P$-wave phases found in the reference amplitude Eq.~\ref{eq_ref_amp}, we exclude the deeply bound level seen in $[000]T_1^-$ and moving frame irreps. This results in 14 energy levels that are used to obtain the amplitudes given in table~\ref{tab:param1}, with only a constant $K$-matrix in $P$-wave. The results given in table~\ref{tab:param2} use the same 20 levels utilised for the reference amplitude, Eq.~\ref{eq_ref_amp}. In the region above threshold, all of the $P$-wave amplitudes produce phase shifts that are approximately zero. 

In the following we present the minima found for a few key parameterisations that are discussed further in section \ref{sec:interpretation}.
Using the $S$-wave Breit-Wigner as defined in Eq.~\ref{eq_BW} (parameterisation (q) of table~\ref{tab:param2}) gives the following parameter values
\begin{equation*}
\begin{aligned}[t]
\begin{matrix}
m_R &= &(0.3913  \pm 0.0041 \pm 0.0014) \cdot a_t^{-1} \\
g_R &= &(5.39    \pm 0.45   \pm 0.11) \\
m_1 &= &(0.33014 \pm 0.00016 \pm 0.00003) \cdot a_t^{-1} \\
g_1 &= &(0.3     \pm 1.3    \pm 0) \\

\end{matrix}
\end{aligned}
\qquad
\begin{aligned}[t]
\begin{bmatrix}
1.00 & 0.92 & 0.26 & -0.03  \\
& 1.00 & 0.17 & -0.04  \\
& & 1.00 & -0.01 \\
& & & 1.00

\end{bmatrix}
\end{aligned}
\end{equation*}
\begin{equation}
\chi^2/N_{\text{dof}} = 14.63 / (20 - 4) =  0.91\,,
\end{equation}
where the subscript ``1'' indicates the $P$-wave parameters, and the others are $S$-wave.

\begin{figure}[tb]
	\centering
		\includegraphics[width=0.75\textwidth]{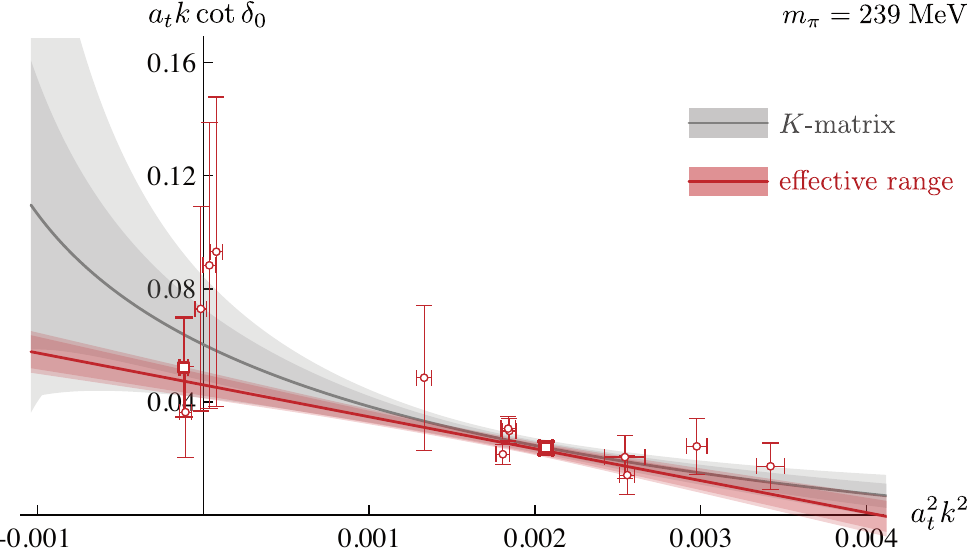}
	\caption{The $S$-wave effective range parameterisation from Eq.~\ref{eq_fit_ere} (red) and reference amplitude Eq.~\ref{eq_ref_amp} (grey) plotted as $a_t k\cot\delta$ as a function of the momentum-squared $a_t^2 k^2$. The discrete points show where the energies constrain the amplitudes by using Eq.~\ref{eq_det} to obtain $t$ level-by-level. The bold square points are obtained from $[000]A_1^+$, and the other points are from moving frame $A_1$ irreps with $P$-wave and higher partial waves fixed to zero. Both amplitudes are determined from the same spectra, as described in the text.}
	\label{fig_860_ere}
\end{figure}

Fitting the effective range parameterisation as defined in Eq.~\ref{eq_ER} (parameterisation (m) of table~\ref{tab:param2}) gives
\begin{equation*}
\begin{aligned}[t]
\begin{matrix}
a_0 &= &(21.9    \pm 1.9 \pm 0.5) \cdot a_t \\
r_0 &= &(-22.1   \pm 4.3 \pm 1.6) \cdot a_t \\
m_1 &= &(0.33013 \pm 0.00016 \pm 0.00003) \cdot a_t^{-1} \\
g_1 &= &(0.2     \pm 1.1 \pm 0.5) \\

\end{matrix}
\end{aligned}
\qquad
\begin{aligned}[t]
\begin{bmatrix}
1.00 & 0.90 & 0.09 & -0.25  \\
& 1.00 & 0.21 & -0.23  \\
& & 1.00 & -0.08 \\
& & & 1.00

\end{bmatrix}
\end{aligned}
\end{equation*}
\begin{equation}
\chi^2/N_{\text{dof}} = 14.81 / (20 - 4) =  0.93\;.
\label{eq_fit_ere}
\end{equation}
We plot this amplitude as $a_tk\cot\delta$ as a function of $a_t^2k^2$ compared to the reference amplitude Eq.~\ref{eq_ref_amp} in Fig.~\ref{fig_860_ere}. This shows the subthreshold constraint from $[000]A_1^+$, $[100]A_1$ and $[110]A_1$. The amplitudes both describe the spectra well. However they do differ within uncertainties, as can be seen in the figure.

For the unitarised chiral amplitude, Eq.~\ref{eq_K_chipt}, (parameterisation (s) of table~\ref{tab:param2}) we obtain
\begin{equation*}
\begin{aligned}[t]
\begin{matrix}
F           &= &(0.0191  \pm 0.0016 \pm 0.0002) \cdot a_t^{-1} \\
\alpha(\mu) &= &(-1.92   \pm 0.25 \pm 0.14) \\
m_1         &= &(0.33020 \pm 0.00016 \pm 0.00003) \cdot a_t^{-1} \\
g_1         &= &(0.76     \pm 0.39    \pm 0.11)  \\

\end{matrix}
\end{aligned}
\qquad
\begin{aligned}[t]
\begin{bmatrix}
1.00 & -0.99 & -0.18 & 0.30  \\
& 1.00 &  0.21 & -0.28  \\
& & 1.00 & -0.12 \\
& & & 1.00
\end{bmatrix}
\end{aligned}
\end{equation*}
\begin{equation}
\chi^2/N_{\text{dof}} = 13.78 / (20 - 4) =  0.86\;.
\end{equation}
The parameter $F$ corresponds to $116\pm 10$~MeV when converted to scale-set units. The definition of $F$ differs by $1/\sqrt{2}$ from that used for $f_\pi$ by the Particle Data Group~\cite{Zyla:2020zbs}.

The amplitudes in tables \ref{tab:param1} and \ref{tab:param2} are very similar at real energies. This is presented in section~\ref{sec:mass_dep} as the outer red bands in the left panel of Fig.~\ref{fig:rhot_mass_dep}. However, they do differ when continued to complex energies, which we investigate in the next section.

\begin{table}
		\begin{tabular}{lllcc}
			\toprule
			{} &                                         $\ell=0$ parameterisation & $\ell=1$ parameterisation & $N_{\text{pars}}$ &  $\chi^2/N_{\mathrm{dof}}$ \\
			\midrule
			\multicolumn{5}{l}{K-matrix with a Chew-Mandelstam $I(s)$ in both partial waves} \\
			(ax) &                                                $K = \frac{g^2}{m^2 - s}$ &        $K = \gamma_1$ &                 3 &              1.12 \\
			(bx) &                                 $K = \frac{g^2}{m^2 - s} + \gamma^{(0)}$ &        $K = \gamma_1$ &                 4 &              1.15 \\
			(cx) &                                $K = \frac{g^2}{m^2 - s} + \gamma^{(1)} \hat{s}$ &        $K = \gamma_1$ &                 4 &              1.15 \\
			(dx) &                                   $K = \frac{(g + g^{(1)}s)^2}{m^2 - s}$ &        $K = \gamma_1$ &                 4 &              1.15 \\
			(ex) &                                            $K^{-1} = c^{(0)} + c^{(1)} \hat{s}$ &        $K = \gamma_1$ &                 3 &              1.12 \\
			(fx) &                           $K^{-1} = \frac{c^{(0)} + c^{(1)} \hat{s}}{c^{(2)} \hat{s}}$ &        $K = \gamma_1$ &                 4 &              1.15 \\
			\midrule
			\multicolumn{5}{l}{K-matrix with $I(s) = -i \rho(s)$ in both partial waves} \\
			(gx) &                                                $K = \frac{g^2}{m^2 - s}$ &        $K = \gamma_1$ &                 3 &              1.13 \\
			(hx) &                                 $K = \frac{g^2}{m^2 - s} + \gamma^{(0)}$ &        $K = \gamma_1$ &                 4 &              1.16 \\
			(ix) &                                $K = \frac{g^2}{m^2 - s} + \gamma^{(1)} \hat{s}$ &        $K = \gamma_1$ &                 4 &              1.19 \\
			(jx) &                                   $K = \frac{(g + g^{(1)}s)^2}{m^2 - s}$ &        $K = \gamma_1$ &                 4 &  $\it 1.37$ \\
			(kx) &                                                $K^{-1} = c^{(0)} + c^{(1)} \hat{s}$ &        $K = \gamma_1$ &                 3 &              1.13 \\
			(lx) &                           $K^{-1} = \frac{c^{(0)} + c^{(1)} \hat{s}}{c^{(2)} \hat{s}}$ &        $K = \gamma_1$ &                 4 &              1.16 \\
			\midrule
			\multicolumn{5}{l}{Effective range} \\
			(mx) &                          $k \cot \delta_0 = 1/a_0 + \frac 1 2 r_0^2 k^2$ &        $K = \gamma_1$ &                 3 &              1.14 \\
			(nx) &             $k \cot \delta_0 = 1/a_0 + \frac 1 2 r_0^2 k^2 + P_{2,0}k^4$ &        $K = \gamma_1$ &                 4 &  $\it 1.12$ \\
			\midrule
			\multicolumn{5}{l}{Breit-Wigner} \\
			(ox) &  $t = \frac 1 {\rho} \frac {m_R \Gamma_0}{m_R^2 - s - i m_R \Gamma_0}$ &        $K = \gamma_1$ &                 3 &              1.13 \\
			\midrule
			\multicolumn{5}{l}{Unitarised $\chi_{\text{PT}}$} \\
			(px) &                                                                       $t^{-1} = \big( - \frac{1}{16 \pi} \mathcal{V}_{J=0} \big)^{-1} + 16\pi G_{\text{DR}}$ &        $K = \gamma_1$ &                 3 &              1.10 \\
			\bottomrule
		\end{tabular}
		\caption{The parameterisations used that excluded the deeply-bound level around $a_tE_\cm=0.33$. $N_{\text{pars}}$ indicates the number of free parameters in each parameterisation. An italicised $\chi^2/N_{\mathrm{dof}}$ value indicates this fit was not included in the amplitude figure and pole values due to an additional pole found on the physical sheet. }
	\label{tab:param1}
\end{table}

\begin{table}[tb]
		\begin{tabular}{lllcc}
			\toprule
			{} &                                                    $\ell=0$ parameterisation &                                                    $\ell=1$ parameterisation & $N_{\text{pars}}$ &  $\chi^2/N_{\mathrm{dof}}$ \\
			\midrule
			\multicolumn{5}{l}{K-matrix with Chew-Mandelstam $I(s)$ in both partial waves} \\
			ref. &                                                $K = \frac{g^2}{m^2 - s} +\gamma^{(0)}$ &                                            $K = \frac{g_1^2}{m_1^2 - s}$ &                 5 &              0.90 \\
			(a) &                                                $K = \frac{g^2}{m^2 - s}$ &                                            $K = \frac{g_1^2}{m_1^2 - s}$ &                 4 &              0.90 \\
			(b) &                                $K = \frac{g^2}{m^2 - s} + \gamma^{(1)} \hat{s}$ &                                            $K = \frac{g_1^2}{m_1^2 - s}$ &                 5 &              0.90 \\
			(c) &                                   $K = \frac{(g + g^{(1)}s)^2}{m^2 - s}$ &                                            $K = \frac{g_1^2}{m_1^2 - s}$ &                 5 &              0.90 \\
			(d) &                                            $K^{-1} = c^{(0)} + c^{(1)} \hat{s}$ &                                            $K = \frac{g_1^2}{m_1^2 - s}$ &                 4 &              0.90 \\
			(e) &                           $K^{-1} = \frac{c^{(0)} + c^{(1)} \hat{s}}{c^{(2)} \hat{s}}$ &                                            $K = \frac{g_1^2}{m_1^2 - s}$ &                 5 &              0.90 \\
			(f) &                 $K = \frac{g^2}{m^2 - s} + \gamma^{(0)} + \gamma^{(1)} \hat{s}$ &                                            $K = \frac{g_1^2}{m_1^2 - s}$ &                 6 &              $\it 0.94^*$ \\
			\midrule
			\multicolumn{5}{l}{K-matrix with $I(s) = -i \rho(s)$ in both partial waves} \\
			(g) &                                 $K = \frac{g^2}{m^2 - s} + \gamma^{(0)}$ &                                            $K = \frac{g_1^2}{m_1^2 - s}$ &                 5 &              0.90 \\
			(h) &                                                $K = \frac{g^2}{m^2 - s}$ &                                            $K = \frac{g_1^2}{m_1^2 - s}$ &                 4 &              0.91 \\
			(i) &                                   $K = \frac{(g + g^{(1)}s)^2}{m^2 - s}$ &                                            $K = \frac{g_1^2}{m_1^2 - s}$ &                 5 &              0.90 \\
			(j) &                                            $K^{-1} = c^{(0)} + c^{(1)} \hat{s}$ &                                            $K = \frac{g_1^2}{m_1^2 - s}$ &                 4 &              0.91 \\
			(k) &                           $K^{-1} = \frac{c^{(0)} + c^{(1)} \hat{s}}{c^{(2)} \hat{s}}$ &                                            $K = \frac{g_1^2}{m_1^2 - s}$ &                 5 &              0.90 \\
			\midrule
			\multicolumn{5}{l}{K-matrix with Chew-Mandelstam $I(s)$ in $S$-wave, Effective range in $P$-wave} \\
			(l) &                                 $K = \frac{g^2}{m^2 - s} + \gamma^{(0)}$ &                          $k \cot \delta_1 = 1/a_1 + \frac 1 2 r_1^2 k^2$ &                 5 &              0.93 \\
			\midrule
			\multicolumn{5}{l}{Effective range in S wave, K-matrix with Chew-Mandelstam $I(s)$ in $P$-wave} \\
		(m) &                          $k \cot \delta_0 = 1/a_0 + \frac 1 2 r_0^2 k^2$ &                                            $K = \frac{g_1^2}{m_1^2 - s}$ &                 4 &              0.93 \\
		(n) &             $k \cot \delta_0 = 1/a_0 + \frac 1 2 r_0^2 k^2 + P_{2,0}k^4$ &                                            $K = \frac{g_1^2}{m_1^2 - s}$ &                 5 &  $\it 0.88^{\dagger}$ \\
			\midrule
			\multicolumn{5}{l}{Effective range in both partial waves} \\
			(o) &                          $k \cot \delta_0 = 1/a_0 + \frac 1 2 r_0^2 k^2$ &                          $k \cot \delta_1 = 1/a_1 + \frac 1 2 r_1^2 k^2$ &                 4 &              0.93 \\
			(p) &             $k \cot \delta_0 = 1/a_0 + \frac 1 2 r_0^2 k^2 + P_{2,0}k^4$ &                          $k \cot \delta_1 = 1/a_1 + \frac 1 2 r_1^2 k^2$ &                 5 &  $\it 0.91^{\dagger}$ \\
			\midrule
			\multicolumn{5}{l}{Breit-Wigner in $S$-wave, K-matrix with Chew-Mandelstam $I(s)$ in $P$-wave} \\
			(q) &  $t = \frac 1 {\rho} \frac {m_R \Gamma_0}{m_R^2 - s - i m_R \Gamma_0}$ &                                            $K = \frac{g_1^2}{m_1^2 - s}$ &                 4 &              0.91\\
			\midrule
			\multicolumn{5}{l}{First-order unitarised $\chi_{\text{PT}}$} \\
			(s) &                                                                       $t^{-1} = \big( - \frac{1}{16 \pi} \mathcal{V}_{J=0} \big)^{-1} + 16\pi G_{\text{DR}}$ &        $K = \frac{g_1^2}{m_1^2 - s}$ &                 4 &              0.86 \\
			\bottomrule
		\end{tabular}\\
		
	$\dagger$ - these amplitudes were found to have physical sheet poles in $S$-wave\\
	$*$ - this amplitude was found to have an additional resonance pole, as described in the text
	
	\caption{The parameterisations used that included the $P$-wave deeply bound level. $N_{\text{pars}}$ indicates the number of free parameters in each parameterisation. An italicised $\chi^2/N_{\mathrm{dof}}$ value indicates this fit was not included in the amplitude figure and pole values, due to the presence of either physical sheet poles, or additional resonance poles close to the left-hand cut.}
	\label{tab:param2}
\end{table}

\section{Scattering amplitude poles}
\label{sec:poles}

In this section we analyse the scattering amplitudes presented above for poles, by analytically continuing to complex $s=E_\cm^2$. The amplitudes have been constrained only at real energies, and when continuing to complex values it is possible that even apparently similar amplitudes differ away from the real axis. However, if a nearby pole is present it is often a universal feature across parameterisations that have similar shapes on the real axis. By extracting the poles of the amplitudes we obtain the essence to compare among different parameterisations, calculations, and experiments.

Scattering amplitude poles unify bound-states and resonances in a single quantity that provides information about the spectral content of the channels under consideration.  In the region of a pole, the $t$-matrix is dominated by a term $t\sim c^2/(s_0-s)$ where $c^2$ is the residue and $s_0$ is the pole position. The factorised pole residue $c$ gives a measure of the coupling to the decay channel.

The amplitudes we have used are analytic in $s$, except for cuts due to the $\cm$ momentum $k(s)$ square-root function and poles. The $s$-channel cut leads to a multi-sheeted complex $s$ plane, where each contributing channel doubles the number of sheets. Sheets can be labelled by the sign of the imaginary part of the momentum $k_i(s)$ for channel $i$. In this analysis we only consider a single channel and therefore the amplitudes as functions of $s$ live on two sheets. The sheet with $\text{sgn}( \text{Im } k ) = -1$ is referred to as the unphysical sheet whereas the one with $\text{sgn}( \text{Im } k ) = +1$ is called the physical sheet. The amplitudes utilised do not incorporate any effects due to exchange processes that introduce additional (``left-hand'') cuts beginning at $a_t\sqrt{s}=0.306$, extending to negative $s$.

Causality restricts complex poles to occur only on the unphysical sheet, and with the amplitudes we consider they will appear as complex-conjugate pairs.  Bound states correspond to poles on the real axis below threshold on the physical sheet. Resonances are found at complex energies, with $\sqrt{s_0}=m-i\Gamma/2$ where $m$ is the mass and $\Gamma$ is the width. 
We begin by investigating the $S$-wave amplitudes for poles.

\subsection{$S$-wave pole}

\begin{figure}[tb]
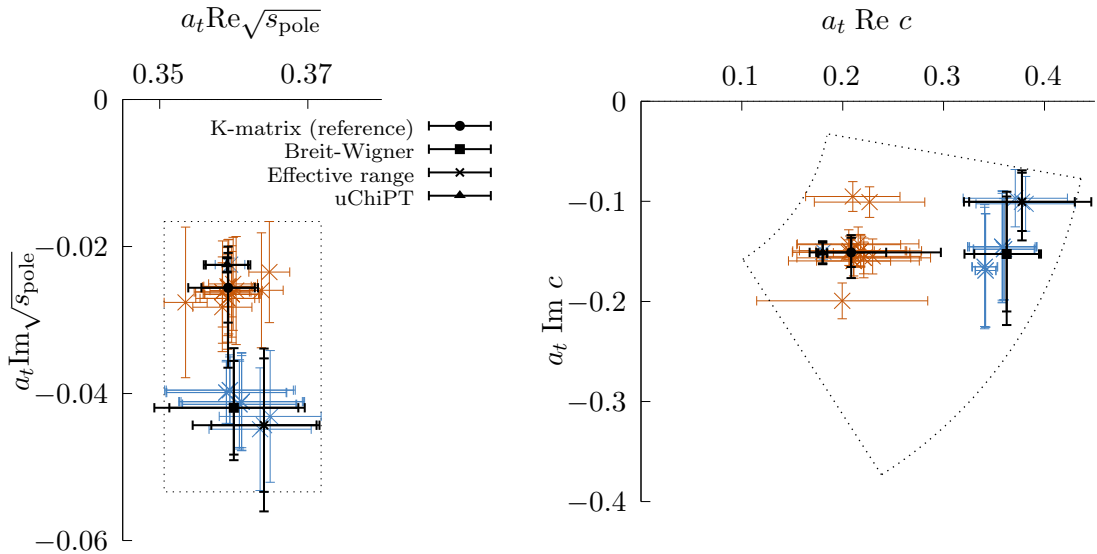

	\begin{subfigure}[t]{0.49\textwidth}
		\hspace*{-3.5cm}
		\graphicspath{{plots/}}
		\input{plots/poles.tex}
	\end{subfigure}
	\begin{subfigure}[t]{0.49\textwidth}
		\graphicspath{{plots/}}
		\input{plots/couplings.tex}
	\end{subfigure}
	\caption{Poles on complex energy plane (left) and couplings (right). The black filled circle corresponds to the reference amplitude. Other amplitudes discussed in the text are shown with different markers (see key). The coloured datapoints in both panels show the spread of poles and couplings produced by the complete set of parameterisation variations (see tables \ref{tab:param1} and \ref{tab:param2}). Orange crosshairs correspond to three-parameter $S$-wave amplitudes, blue crosshairs to two-parameter ones. The dotted rectangle encompasses the entire spread of the parameterisations including their statistical uncertainties, but excluding variations of mass or anisotropy for amplitudes other than the reference parameterisation.}
	\label{fig_S_poles}
\end{figure}

The reference amplitude, Eq.~\ref{eq_ref_amp}, has an $S$-wave pole on the unphysical sheet at $a_t\sqrt{s_0} = (0.3592 \pm 0.0036) - \frac{i}{2}(0.0512 \pm 0.0095)$, shown as the filled black circle in the left panel of Fig.~\ref{fig_S_poles}. The amplitude rises rapidly from threshold, and this feature corresponds to a pole in all parameterisations, shown in Fig.~\ref{fig_S_poles}.

Considering all the parameterisations, the $S$-wave poles form two clusters, one with $-2a_t\mathrm{Im}\sqrt{s_0}\approx 0.05$ (orange markers in Fig.~\ref{fig_S_poles}), and one slightly deeper in the complex plane with $-2a_t\mathrm{Im}\sqrt{s_0}\approx 0.08$ (blue). While both clusters correspond to amplitudes with perfectly acceptable $\chi^2/N_{\mathrm{dof}}$ values, the nearer cluster corresponds to three-parameter $S$-wave amplitudes, and the deeper cluster arises from amplitudes with two free parameters. In table~\ref{tab_Kmat_npars}, we compare 2, 3, and 4 parameter $S$-wave $K$-matrix fits. The 3-parameter fit corresponds to the reference parameterisation, Eq.~\ref{eq_ref_amp}. The two-parameter fit results in a deeper pole (amplitude (a)). The four-parameter fit with a linear term $\gamma^{(1)}\hat{s}$ (amplitude (f)) results in an amplitude with two poles, one around $a_tm\approx 0.29$, far below threshold but close to the left-hand cut, and one similar to those found for the two and three-parameter fits. The $\chi^2/N_\mathrm{dof}$ increases suggesting that there is too much freedom. We choose to exclude parameterisations such as this that produce poles in the energy region of the left cut. However these all have a pole consistent with the dotted region marked in Fig.~\ref{fig_S_poles}, and so this choice does not affect the final result. We also exclude any parameterisation that produces physical sheet poles in $S$-wave. 

In table~\ref{tab_Kmat_npars}, the magnitude of the pole residue correlates with the magnitude of the imaginary part. This suggests how these amplitudes achieve very similar behaviours at real energies despite having slightly different pole positions. Although the data constrain the real part of the pole position relatively well, some freedom remains in the imaginary part. Nevertheless, the pole is present in all parameterisations and so appears to be a necessary feature to describe the lattice QCD spectra.

\begin{table}
\centering
\begin{center}
\addtolength{\tabcolsep}{-3pt}
\begin{tabular}{ccccc|c|ccc}
amp & $a_tm$ & $a_tg$ & $\gamma^{(0)}$ & $\gamma^{(1)}$ & $\frac{\chi^2}{N_\mathrm{dof}}$ & $\mathrm{Re}(a_t\sqrt{s_0})$ & -$2\mathrm{Im}(a_t\sqrt{s_0})$ & $a_t|c|$\\[0.4ex]
\hline
(a) & 0.3916(42) & 0.313(22) & -        & -         & 0.90 & 0.3590(80) & 0.0797(83) & 0.381(33)\\
ref.& 0.4011(98) & 0.419(83) & -2.0(13) & -         & 0.90 & 0.3592(35) & 0.0512(95) & 0.257(33)\\
(f) & 0.4222(92) & 0.789(57) & -8.6(16) & -14.7(87) & {\it 0.94} & {\it 0.3638(35)} & {\it 0.0465(74)} & {\it 0.218(27)}
\end{tabular}
\addtolength{\tabcolsep}{3pt}
\end{center}
\caption{The result of varying the number of free parameters in the $S$-wave amplitude with a two-parameter $P$-wave as used in Eq.~\ref{eq_ref_amp}. ``ref.'' indicates the reference amplitude, Eq.~\ref{eq_ref_amp}. The final amplitude (f) results in a parameterisation that produces two poles, one of them around $a_tm\approx 0.29$, far below threshold. The italics highlight that this amplitude contains this lower resonance pole.}
\label{tab_Kmat_npars}
\end{table}

Among the parameterisation variations we have implemented is a unitarised chiral amplitude, as mentioned above and described in detail in appendix~\ref{app:sec:chipt}. This amplitude also produces a pole that lies within the cluster of three-parameter $S$-wave amplitudes and is marked by a black triangle in Fig.~\ref{fig_S_poles}. The Breit-Wigner is also indicated (black square), which results in a pole with a larger imaginary part than many of the other amplitudes.

The data demands a pole is present, however the scatter of pole positions shown in Fig.~\ref{fig_S_poles} demonstrates that using any single parameterisation does not necessarily give a reliable estimate of the uncertainties. Our final value for the pole position and coupling taking into account the statistical uncertainty from all parameterisations is 
\begin{align}
a_t\sqrt{s_0} &= (0.361 \pm 0.011) - \tfrac{i}{2} (0.070 \pm 0.037) \\
a_t c         &= ( 0.32 \pm 0.13 ) \exp i\pi( -0.59 \pm 0.41 ) \,,
\end{align}
this corresponds to the dotted area in Fig.~\ref{fig_S_poles}. In physical units this corresponds to
\begin{align}
\sqrt{s_0} &= \left((2196 \pm 64) - \tfrac{i}{2} (425 \pm 224)\right) \:\mathrm{MeV} \\
 c         &= \left((1916 \pm 776)\exp i\pi( -0.59 \pm 0.41 )\right) \:\mathrm{MeV}\;.
\end{align}

The amplitudes were also investigated for additional poles at higher energies. However, none were consistently found across the many parameterisations. This indicates that the energy levels determined in this elastic energy region only demand the presence of a single resonance pole. We do not rule out the possibility of any additional poles beyond where the amplitudes have been constrained; such additional poles have been suggested, for example in ref.~\cite{Albaladejo:2016lbb}.

\subsection{$P$-wave pole}

In $J^P=1^-$ a deeply-bound $D^\star$ pole is found at a similar energy to the energy level far below threshold in all irreps where $J^P=1^-$ subduces. This pole does not appear to significantly influence the physical scattering region as can be seen for example in Fig.~\ref{SP_phase_shift}, and although from our amplitudes in table~\ref{tab:param2} a pole coupling can be extracted, the uncertainties are very large and the coupling does not appear to be particularly meaningful.\footnote{We have verified that the residue of the pole has the appropriate sign for a bound state.} 

The mass found is consistent with the result considering only $q\bar{q}$ operators in table~\ref{table:mesons}, suggesting little influence on this deeply-bound state from the $D\pi$ operators. The experimental $D^\star$ is observed to be very narrow, found close to $D\pi$ threshold. With the $P$-wave phase space opening relatively slowly, it would likely require close-to-physical pion masses to observe significant shifts away from the non-interacting $D\pi$ energies, and thus determine a coupling to the $D\pi$ decay channel.

Across all parameterisations that include a $P$-wave pole term (see table \ref{tab:param2}) we obtain
\begin{align}
a_t\sqrt{s_0} &= 0.3301 \pm 0.0012\,.
\end{align}
This is consistent with the deeply bound state seen in irreps where $J^P=1^-$ appears, as shown in Figs.~\ref{fig:spec1} and \ref{fig:spec2}. In physical units, the pole is located at
\begin{align}
\sqrt{s_0} &= \left(2006.9 \pm 7.4\right)\:\mathrm{MeV} \;.
\end{align}

\section{Interpretation}
\label{sec:interpretation}

We now interpret our results, first comparing to earlier work at a light quark mass corresponding to a larger pion mass, before considering the composition of the scalar state. We also compare with studies of $DK$ scattering at the same pion masses in the context of $SU(3)$ flavour symmetry.

\subsection{Light quark mass dependence}
\label{sec:mass_dep}

\begin{figure}[tb]
\begin{center}
\includegraphics[width=0.95\textwidth]{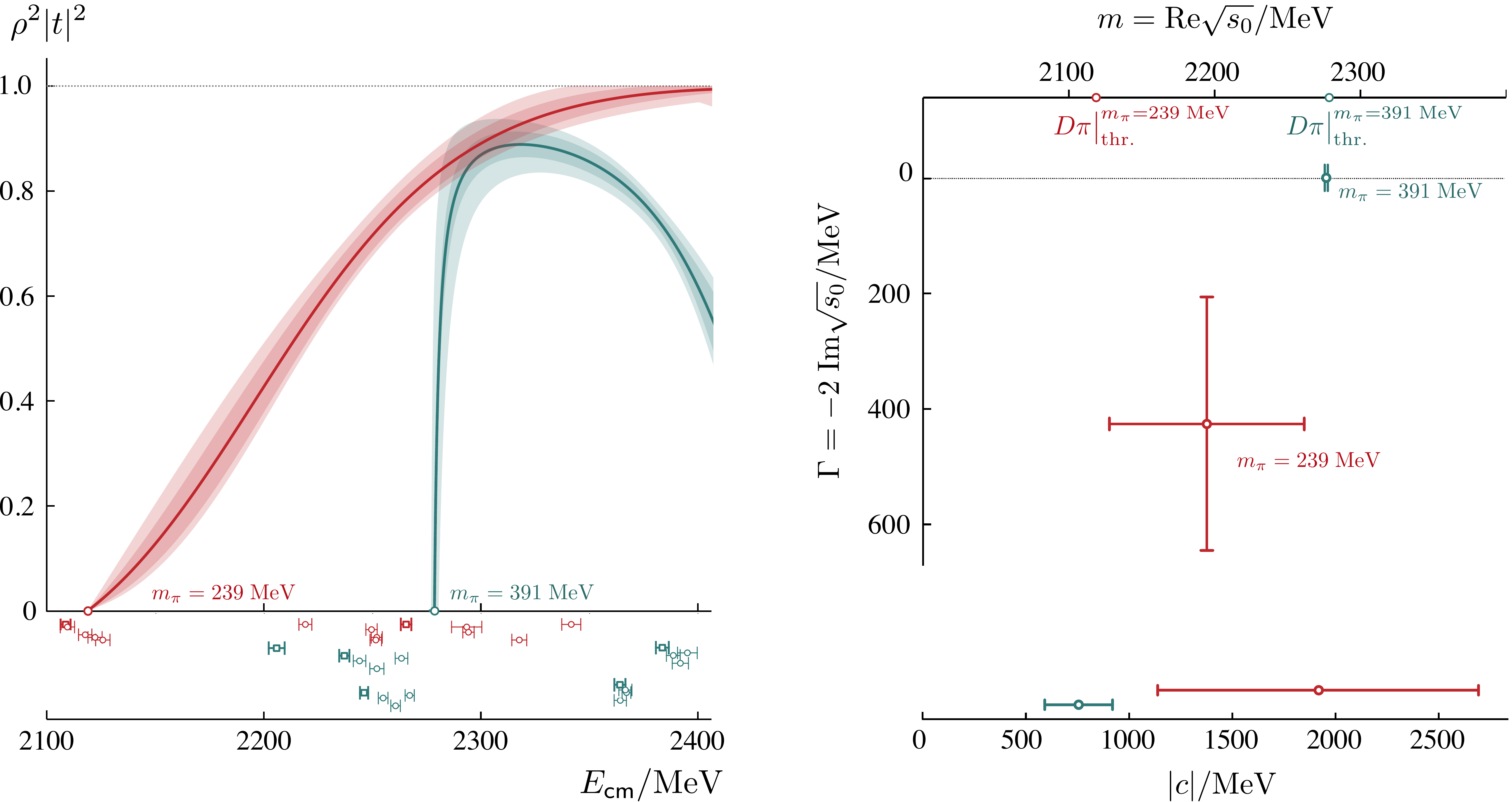}
\caption{The left panel shows the reference $S$-wave scattering amplitude at $m_\pi=239$~MeV (red) and 391~MeV (blue) plotted as $\rho^2|t|^2$ with the energies that were used to constrain them shown below. The bold square points are from $[000]A_1^+$, the other points are from moving frame irreps. The inner bands show the statistical uncertainty from the $\chi^2$ minimisation. The outer band for $m_\pi=239$~MeV includes variation over mass, anisotropy and parameterisations. The outer band for $m_\pi=391$~MeV includes variation over only mass and anisotropy; Ref.~\cite{Moir:2016srx} found only a small effect from varying the parameterisation for this elastic system with a near-threshold bound state. The upper right panel shows the $S$-wave pole positions including the additional uncertainty found from the variation over parameterisation, which is significant for $m_\pi=239$~MeV. The pole at the lower pion mass is a resonance found on the unphysical sheet, and at the higher pion mass is a bound state found on the physical sheet. The lower right panel shows the magnitudes of pole couplings to the $S$-wave $D\pi$ channel.}
\label{fig:rhot_mass_dep}
\end{center}
\end{figure}

\begin{figure}[tb]
\begin{center}
\includegraphics[width=0.8\textwidth]{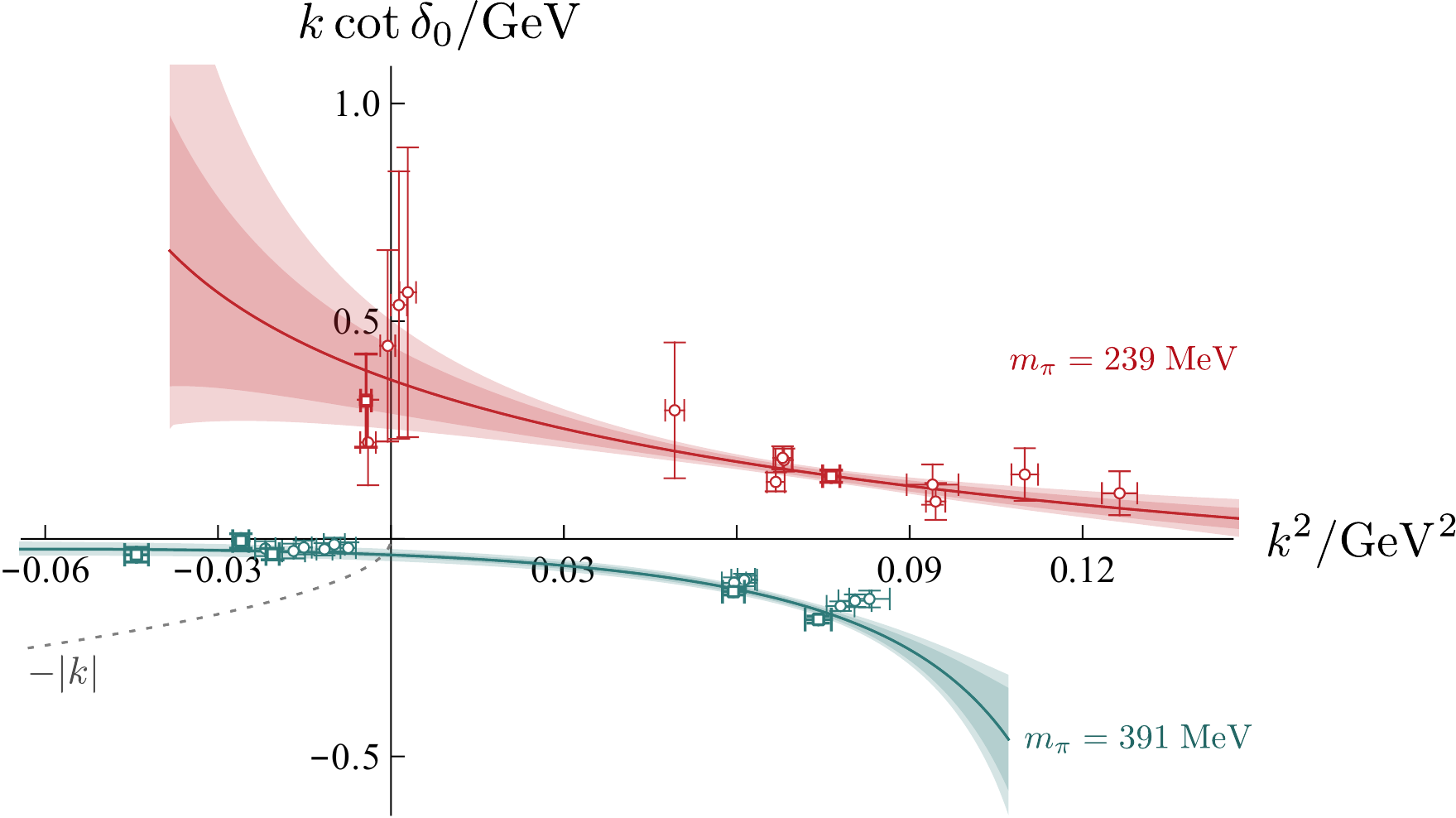}
\caption{The $S$-wave scattering amplitudes at $m_\pi=239$ and 391 MeV plotted as $k\cot\delta$ as a function of $k^2$ in scale-set units. Red shows the $m_\pi=239$~MeV amplitude and the $m_\pi=391$ MeV amplitude is shown in blue. The points shown come from using the finite volume energies individually in the L\"uscher determinant condition. The bold square points are obtained from irreps at rest, the other points are obtained from moving frame irreps. In both cases the $P$-wave does not have a significant impact and was fixed to zero where it appears in moving frames. The bound state mass at $m_\pi=391$~MeV can be read off from the intersection of the blue curve with the dotted $-|k|$ curve at negative $k^2$.}
\label{fig:kcot}
\end{center}
\end{figure}

\begin{figure}[tb]
\begin{center}
\includegraphics[width=0.99\columnwidth]{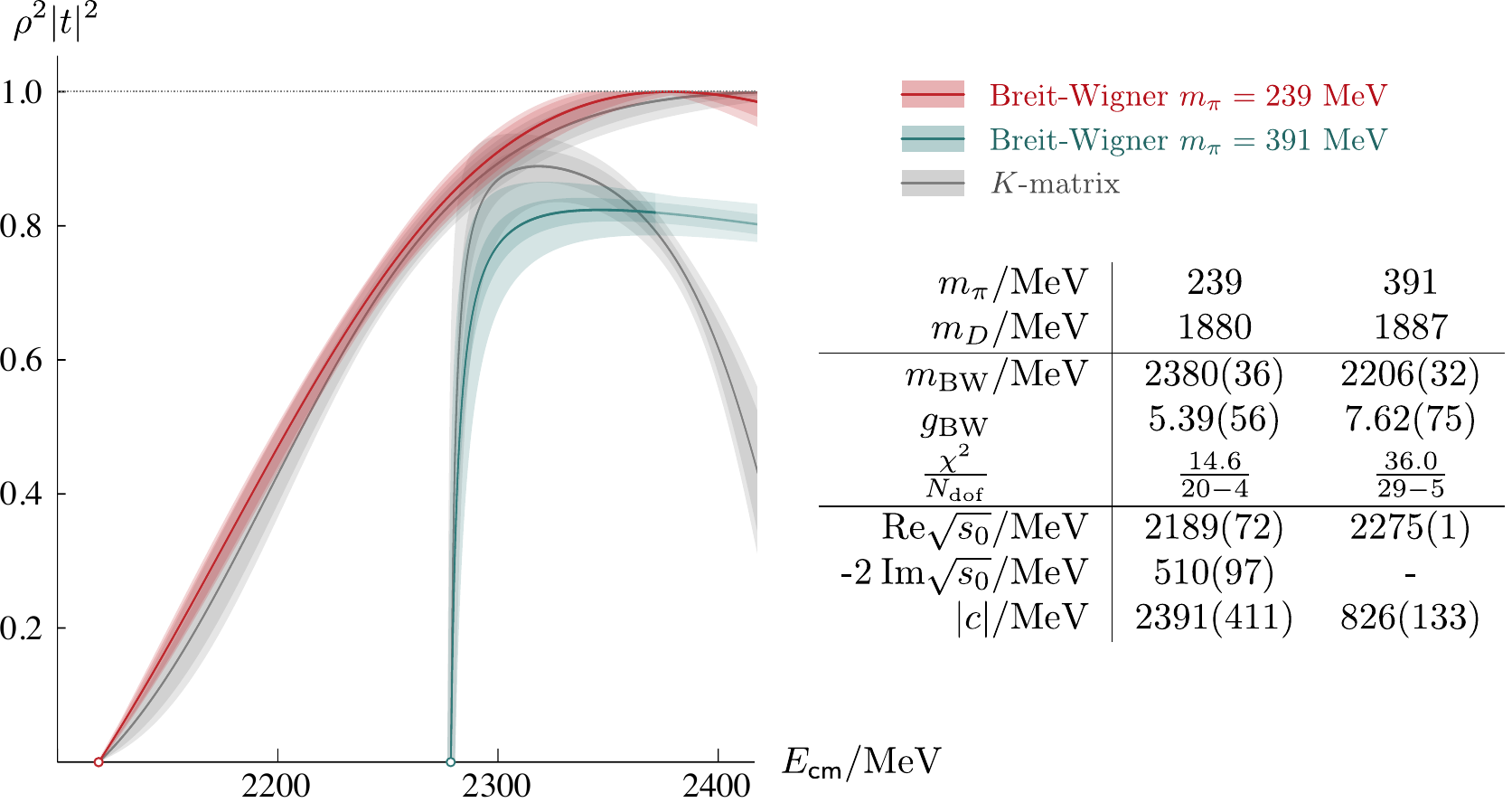}
\end{center}
\caption{$S$-wave Breit-Wigner parameterisations compared to the reference $K$-matrix fits at both light quark masses - the numerical uncertainties shown include the statistical, and the mass, anisotropy variations summed together. The complex $t$-matrix pole position is indicated with mass $m=\mathrm{Re}\sqrt{s_0}$, and width $\Gamma= - 2\,\mathrm{Im}\sqrt{s_0}$. The pole coupling $|c|$ to $D\pi$ is also indicated. The fit range used for the Breit-Wigner fit is shortened relative to the $K$-matrix at $m_\pi=391$~MeV as described in the text and indicated in the plot by the faded region at the highest energies. }
\label{fig_BW_compared} 
\end{figure}

In Ref.~\cite{Moir:2016srx}, $D\pi$ scattering was studied on three volumes with a light quark mass corresponding to $m_\pi=391$~MeV. A near-threshold bound state was found in $S$-wave with a strong coupling to the $D\pi$ channel, that influenced a broad energy region. The reference amplitude from that work\footnote{Eq.~3.3 of Ref.~\cite{Moir:2016srx}.}, updated with an improved estimate of the $D$-meson mass from Ref.~\cite{Cheung:2020mql}, results in
\begin{center}
\begin{tabular}{rll}
$m \;\;=$                   & $\;\;(0.3866 \pm 0.0026 \pm 0.0009) \cdot a_t^{-1}$   & 
\multirow{6}{*}{ $\begin{bmatrix} 1.00 & -0.67 &  -0.39  &  0.28  & -0.06   &  0.10 \\ 
                                    &  1.00    &   0.94  &  0.03  & -0.09   &  0.09 \\
                                    &          &   1.00  &  0.16  & -0.14   &  0.13 \\   
                                    &          &         &  1.00  & -0.76   & -0.38 \\
                                    &          &         &        &  1.00   &  0.56 \\
                                    &          &         &        &         &  1.00    \end{bmatrix}$ } \\
$g \;\;=$                   & $\quad\;\; (0.73 \pm 0.07 \pm 0.16) \cdot a_t^{-1}$   & \\
$\gamma \;\;= $             & $\quad\;\;\; 13.4 \pm 3.0 \pm 4.2 $   & \\
$m_1 \;\;=$                 & $(0.35445 \pm 0.00017 \pm 0.00002) \cdot a_t^{-1}$   & \\
$g_1 \;\;=$                 & $\quad\quad 1.49 \pm 0.34 \pm 0.01$    & \\
$\gamma_1 \;\;= $           & $\quad(-107 \pm 43 \pm 9) \cdot a_t^{2} $   & \\[1.3ex]
&\multicolumn{2}{l}{ $\chi^2/ N_\mathrm{dof} = \frac{41.44}{33-6} = 1.53 $\,,}
\end{tabular}
\end{center}
\vspace{-0.8cm}
\begin{equation} \label{eq_fit_840_SP}\end{equation}\\
and the corresponding amplitude is plotted as the blue curve in Figs.~\ref{fig:rhot_mass_dep} and~\ref{fig:kcot}. This amplitude contains an $S$-wave bound state pole at $a_t\sqrt{s_0}=0.40170(11)(15)$ with a coupling $a_tc_{D\pi}=0.134(9)(29)$.\footnote{Where the meaning of the first and second uncertainties is as defined below Eq.~\ref{eq_PmPp}.} This is broadly in agreement with Ref.~\cite{Moir:2016srx}, with a slightly larger pole coupling. In scale-set units, this bound state remains approximately $2\pm 1$ MeV below threshold, as reported in Ref.~\cite{Moir:2016srx}.

Using instead the Breit-Wigner parameterisation in $S$-wave and the same parameterisation in $P$-wave as in Eq.~\ref{eq_fit_840_SP} results in a poor $\chi^2/N_\mathrm{dof}$. However, reducing the energy region slightly to remove the 4 highest energy points, we find $a_t m_{\mathrm{BW}}=0.3893(23)(33)$, $g_{\mathrm{BW}}=7.62(31)(44)$ and $\chi^2/N_\mathrm{dof}=\frac{36.0}{29-5}=1.5$. This results in a bound-state pole at $a_t\sqrt{s_0}=0.40146(15)(11)$ and coupling $a_tc=0.146(9)(11)$. In Fig.~\ref{fig_BW_compared} we present a comparison of the Breit-Wigner amplitudes fitted at both light quark masses with the reference amplitudes. Due in part to the very large coupling of this state to the $D\pi$ channel, in both cases the Breit-Wigner mass parameter bears little connection to the mass found from the complex pole. At $m_\pi=239$ MeV, the Breit-Wigner produces a broad pole that lies in the cluster of other two-parameter $S$-wave fits, as indicated in Fig.~\ref{fig_S_poles}.

In Fig.~\ref{fig:kcot} we plot the $K$-matrix amplitudes in Eq.~\ref{eq_ref_amp} and Eq.~\ref{eq_fit_840_SP} as $k\cot\delta$ as a function of $k^2$. This is the quantity in which the effective range expansion is usually expressed, $k\cot\delta=\tfrac{1}{a}+\tfrac{1}{2}rk^2+...\,$. In both cases the scattering length has a large magnitude. At the larger pion mass $a<0$, corresponding to a bound-state, seen as the intersection of the $k\cot\delta$ curve and $-|k|$. The amplitude obtained at the smaller pion mass has $a>0$ indicating attraction but no binding; in this case we know that this attraction is due at least in part to the nearby resonance pole. 

A similar behaviour was also observed for $I=0$, $J=0$ $\pi\pi$ scattering~\cite{Briceno:2016mjc}, where a $\sigma$-like pole was found at the same masses, with a near-threshold bound state at the higher mass that evolves into a resonance pole at the lower mass. This similarity is striking when plotted as $k\cot\delta$: Fig.~\ref{fig:kcot} is remarkably similar to the lower panel of Fig.~4 in Ref.~\cite{Briceno:2016mjc}. Conversely, $S$-wave $\pi K$ in $I=1/2$ shows weaker attraction at threshold, perhaps due to a more distant pole~\cite{Pelaez:2020gnd,Danilkin:2020pak}. $S$-wave $DK$ in $I=0$ is found to have bound states at both pion masses~\cite{Cheung:2020mql}.

\begin{figure}[tb]
\centering
\includegraphics[width=0.75\textwidth]{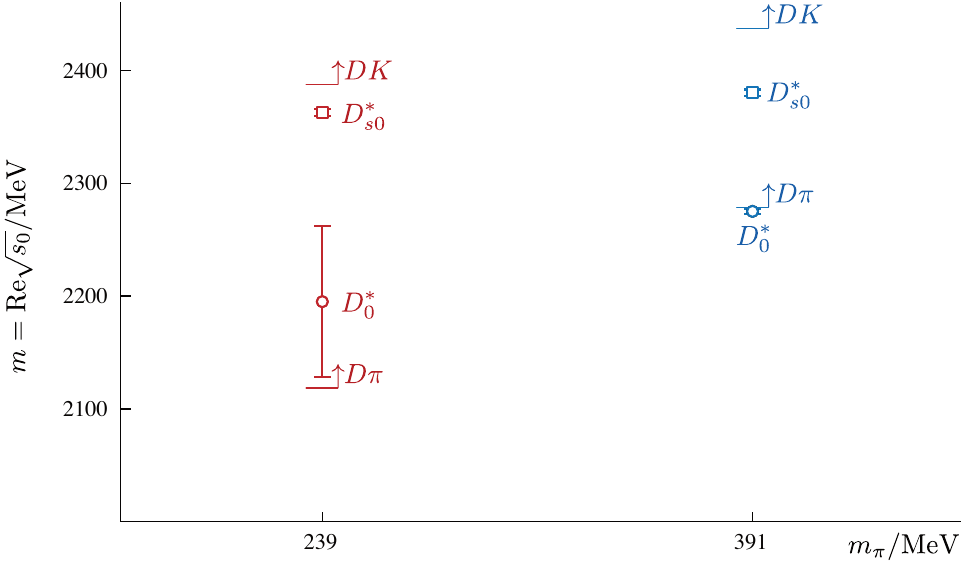}
\caption{A summary of the real parts of the pole positions found in this analysis and Ref.~\cite{Moir:2016srx} in $S$-wave $D\pi$ scattering, and Ref.~\cite{Cheung:2020mql} in $S$-wave $DK$ scattering. The $D_0^\star$ resonance pole found in this study lies on the unphysical sheet with a large width. The other 3 poles are bound states. }
\label{fig:pole_summary}
\end{figure}

One of the initial experimental surprises between the lightest $D_0^\star$ resonance with $c\bar{l}$ quark content, and the lightest $D_{s0}^\star$ resonance with $c\bar{s}$ quark content, was that the $D_{s0}^\star$ was found at a similar mass to the $D_0^\star$. The experimental $D_0^\star$ was also found to be very broad while the $D_{s0}^\star$ was found to be narrow. At $m_\pi=391$~MeV both states appear as bound states below their respective decay channels with the $D_0^\star$ lower in mass than the $D_{s0}^\star$~\cite{Moir:2016srx,Cheung:2020mql}. At $m_\pi=239$~MeV, the $D_0^\star$ pole migrates deep into the complex plane while its real part stays close to $D\pi$ threshold. Nevertheless, the ``natural'' mass ordering with the $c\bar{l}$ being lighter than the $c\bar{s}$ is retained. It should be noted that while in the $m_\pi=391$~MeV calculation, the $D_{s0}^\star$ is more bound than in experiment, for $m_\pi=239$~MeV, the $D_{s0}^\star$ pole is found closer to threshold than it is in experiment~\cite{Cheung:2020mql}. We summarise the real parts of the pole positions as a function of $m_\pi$ in Fig.~\ref{fig:pole_summary}. 

The $D_0^\star$ computed in this study is below the reported mass for the experimentally observed $D_0^\star(2300)$, even though the light quark mass is larger than physical. This was also true in Ref.~\cite{Moir:2016srx}, which at a light quark mass corresponding to a higher pion mass found a bound-state just below $D\pi$ threshold. Based on these two points, the real part of the complex resonance pole appears to move slowly with light quark mass, being found $(77\pm64)$ MeV above threshold at $m_\pi=239$ MeV and $(2\pm1)$ MeV below threshold at $m_\pi=391$ MeV. If this trend continues towards the physical light quark mass, then the current estimate of the $D_0^\star$ mass from the experimental data appears a little too high. However given the large width it is possible that the experimental amplitudes are also compatible with a lower pole mass. Early suggestions this may be the case appeared in Refs.~\cite{vanBeveren:2003kd,vanBeveren:2006st}. During the preparation of this article, Ref.~\cite{Du:2020pui} appeared with a similar conclusion comparing unitarised chiral amplitudes and LHCb data~\cite{Aaij:2016fma}. 

In this study, we have not presented any amplitudes describing the coupled-channel region although a few energy levels were computed that extend above the elastic region. In $[000]A_1^+$ two levels were found coincident with the $D\eta$ and $D_s\bar{K}$ non-interacting levels and we concluded there were not sufficiently many energy levels to make a reliable statement based on this single volume alone. Additional effects may be present from three-body channels such as $D\pi\pi$ and $D^\star\pi\pi$ which is very close to $D_s\bar{K}$ with $m_\pi=239$~MeV. In Ref.~\cite{Moir:2016srx}, some of the levels around $D\eta$ and $D_s\bar{K}$ showed only small shifts, qualitatively similar to what is seen here. However significant effects were observed in the scattering amplitudes.

\subsection{Composition}
The lightest $D_0^\ast$ scalar resonance determined here is significantly below where quark potential models predict the lightest $D_0^\ast$ state to arise~\cite{Godfrey:1985xj}, prompting suggestions of contributions beyond those in the quark model. While at present there is no completely rigorous method to distinguish between different components, models with large contributions from meson-meson (molecular) and compact tetraquark components in addition to quark-model $q\bar{q}$ components are proposed as possible explanations. Here we consider some  measures based upon the pole coupling and proximity to threshold suggested in the literature, and we examine the overlaps with operators resembling $q\bar{q}$-like and $D\pi$-like constructions used to extract the finite volume spectrum.

One measure that is often claimed to distinguish between a compact or a composite state for near-threshold $S$-wave bound-states uses the field renormalisation constant $Z$~\cite{Weinberg:1965zz}, where $Z\to0$ corresponds to a dominant molecular component and $Z\to1$ corresponds to a negligible molecular contribution. This was originally applied to the deuteron indicating that it has a composite nature. Proximity to threshold is the key feature and for states very close to threshold, like the $D_0^\ast$ found at $m_\pi=391$ MeV, the large scattering length $a$ required for a near-threshold state inevitably leads to the suggestion of a large molecular component. Fitting an effective range parameterisation using the same 29 energy levels used for the Breit-Wigner results in $a=-116(15)\:a_t$ and $r=-12.3(8)\:a_t$ and $\tfrac{\chi^2 }{N_\mathrm{dof}}=\tfrac{35.7}{29-5}$. Using $Z=1-\sqrt{a/(a+2r)}$, this corresponds to $Z\approx0.09(8)$ for the bound-state found when $m_\pi=391$~MeV.

Several generalisations of this proposal exist for unstable states~\cite{Morgan:1992ge,Morgan:1993td,Baru:2003qq,Aceti:2012dd,Guo:2015daa,Hyodo:2013iga,Matuschek:2020gqe}, such as 
\begin{align}
|X|&=|1-Z|\nonumber\\
   &=\Biggl|c^2 \frac{d I(s)}{ds}\Big|_{s\to s_0}\Biggr|
\end{align}
from Eq.~(109) of Ref.~\cite{Aceti:2012dd} or Eq.~(25) of Ref.~\cite{Guo:2015daa}, adapted to the definitions used here, where $c^2$ is the pole residue, $s_0$ is the pole position and $I(s)$ is the threshold-subtracted Chew-Mandelstam function. Similar to the case above for bound states, proximity to threshold determines much of the outcome since $I(s)$ moves rapidly around threshold and tends to a constant far away. For the reference parameterisation Eq.~\ref{eq_ref_amp}, this results in $|X|\approx 0.57$, suggesting approximately equal importance of both a compact $q\bar{q}$ contribution and a composite meson-meson contribution. Ref.~\cite{Matuschek:2020gqe} suggests an alternative expression in terms of the effective range formula that results in a similar value. 

An additional place where qualitative information arises in this calculation is in the operator overlaps shown in Fig.~\ref{fig:basis}. The operators used resemble $q\bar{q}$ and two-meson constructions. The $q\bar{q}$-like constructions alone produce a single level within the width of the resonance, similar to as was found in the case of the $\rho$ resonance in Refs.~\cite{Dudek:2012xn,Wilson:2015dqa}. Using only $D\pi$ operators does not produce a single level that is consistent with the spectrum when using all of the operators computed. Adding the $q\bar{q}$ operators to the $D\pi$-like constructions results in a consistent spectrum to that found when using all of the operators in the energy region of the $D_0^\star$. 
Furthermore, these energy levels receive contributions from both types of operators, as shown in Fig.~\ref{fig:basis}.
This is suggestive that both $q\bar{q}$ and $D\pi$ components are necessary since both the $q\bar{q}$-like and $D\pi$-like operators appear essential to determine a reliable spectrum.

\subsection{SU(3) flavour symmetry}

A useful perspective can be gained by considering the limit where the light and strange quark masses are set equal. An SU(3) flavour symmetry then arises between the $\pi$, $K$ and part of the $\eta$ that form the members of an SU(3) octet. $D\pi$ and $DK$ scattering are then closely related, as described in Section 6.4 of Ref.~\cite{Cheung:2020mql}. One key expectation is that the number of poles in each channel should not change as a function of the amount of SU(3) breaking. The pole couplings must be similarly related in the SU(3) limit, and it is possible that these only vary slowly with quark mass.

When comparing the $I=1/2$ $D\pi$ and $I=0$ $DK$ amplitudes at $m_\pi=391$~MeV, SU(3) symmetry is broken but only by a small amount ($m_\pi/m_K\approx0.71$). The closely-related lightest $I=1/2$ $D_0^\star$ and $I=0$ $D_{s0}^\star$ states are both near-threshold bound states, with the $D_{s0}^\star$ being more deeply bound than the $D_0^\star$. In this calculation, working at $m_\pi=239$~MeV, SU(3) is more broken ($m_\pi/m_K\approx0.47$) and the lightest states differ significantly. The ($c\bar{s}$) $D_{s0}^\star$ remains bound, although less than with the heavier light quark, since the scattering $D$ and $K$ mesons both contain light quarks. As we have seen, the ($c\bar{l}$) $D_0^\star$ becomes a resonance with a pole deep in the complex plane.

The poles are found to be strongly-coupled to the relevant meson-meson channels. The $D_{s0}^\star$ at both masses and the resonant $D_0^\star$ studied here have couplings consistent with $|c|\approx 1600$~MeV. The bound $D_0^\star$ found at the larger pion mass has a slightly smaller coupling. However it is found incredibly close to $D\pi$ threshold and this may have an effect as the pole transitions from a bound state to resonance. 

The results obtained here and in Refs.~\cite{Moir:2016srx,Cheung:2020mql} are qualitatively consistent with the expectation that these poles and couplings only change slowly as a function of the amount of $SU(3)$ breaking. 

\section{Summary}
\label{sec:summary}

In this analysis we have presented a computation of the $D_0^\star$ resonance pole from lattice QCD. Working at $m_\pi = 239$~MeV on a single lattice with a spatial volume of approximately $(3.6\;\mathrm{fm})^3$, 20 energy levels were obtained that determine the energy dependence of the infinite volume scattering amplitudes. Significant deviations are observed from the spectrum expected in the absence of interactions, that correspond to a rapid increase in the $S$-wave $D\pi$ scattering amplitude near threshold. Several amplitude parameterisations are considered, and these all produce a single nearby resonance pole with a mass $m=(2196 \pm 64)$ MeV and a width $\Gamma = (425 \pm 224)$ MeV, only $(77\pm64)$ MeV above $D\pi$ threshold, but deep in the complex plane. Much of the uncertainty on the width arises from considering a range of parameterisations.

This calculation has been performed at a heavier-than-physical pion mass, and along with an earlier calculation that found a bound state at a heavier mass~\cite{Moir:2016srx}, likely indicates a slow decrease in the $D_0^\star$ pole mass with decreasing pion mass. At both pion masses, the $D_0^\star$ pole is found to couple strongly to the $D\pi$ channel. At $m_\pi=391$~MeV, the pole is bound but its effect is felt over a broad energy region. At $m_\pi=239$~MeV, a similar large coupling is found but the pole migrates deep into the complex plane, with a significant influence over the whole $S$-wave elastic scattering energy region. The $D_0^\star$ pole computed here has a smaller mass than the currently reported experimental values for the lightest $D_0^\star$ resonance~\cite{Zyla:2020zbs}. Given the large width, the experimentally-determined amplitudes may also be compatible with a smaller mass~\cite{Du:2020pui,Aaij:2016fma}. 

The $D_{s0}^\star$ was also computed at both light quark masses and was found to be significantly heavier than the $D_0^\star$ in both cases~\cite{Cheung:2020mql}. The pole couplings to the relevant thresholds are all found to be large, which may account for some of the puzzling differences highlighted by early experimental studies of these systems. In particular, the difference in widths between the $D_{s0}^\star$ and $D_0^\star$ can be understood from the vastly different phase space available in each case, since the couplings computed are similar. The poles themselves are found to have a ``natural'' ordering with the $c\bar{l}$ lighter than the $c\bar{s}$. 

This calculation completes a quartet of analyses of the $D_0^\star$ and $D_{s0}^\star$ systems at two light quark masses. The puzzle of a broad $D_0^\star$ more massive than the narrow $D_{s0}^\star$ found in experiment is not present for light quark masses corresponding to $m_\pi=239$~MeV and 391~MeV where the $D_0^\star$ pole is found consistently lower in mass than the $D_{s0}^\star$. Furthermore, the couplings of the $D_0^\star$ and $D_{s0}^\star$ poles to $D\pi$ and $DK$ respectively are compatible, suggestive of a common origin. Using a first-principles approach to QCD, with external inputs only to fix quark masses, these analyses thus point to a possible resolution of the puzzling experimental masses and widths.

\bigskip

\begin{acknowledgments}
We would like to acknowledge the contribution by our friend and colleague, David Tims, who sadly passed away during the preparation of this manuscript.
We thank our colleagues within the Hadron Spectrum Collaboration (www.hadspec.org), in particular Jozef Dudek for helpful comments.
DJW acknowledges support from a Royal Society University Research Fellowship. LG acknowledges funding from an Irish Research Council Government of Ireland Postgraduate Scholarship. CET and DJW acknowledge support from the U.K. Science and Technology Facilities Council (STFC) [grant number ST/T000694/1].

The software codes {\tt Chroma}~\cite{Edwards:2004sx}, {\tt QUDA}~\cite{Clark:2009wm,Babich:2010mu}, {\tt QPhiX}~\cite{Joo:2013lwm}, and {\tt QOPQDP}~\cite{Osborn:2010mb,Babich:2010qb} were used to compute the propagators required for this project.
This work used the Wilkes GPU cluster at the University of Cambridge High Performance Computing Service (www.hpc.cam.ac.uk), provided by Dell Inc., NVIDIA and Mellanox, and part funded by STFC with industrial sponsorship from Rolls Royce and Mitsubishi Heavy Industries.
The contractions and additional propagators were also computed on clusters at Jefferson Laboratory under the USQCD Initiative and the LQCD ARRA project.
This research was supported in part under an ALCC award, and used resources of the Oak Ridge Leadership Computing Facility at the Oak Ridge National Laboratory, which is supported by the Office of Science of the U.S. Department of Energy under Contract No. DE-AC05-00OR22725. This research is also part of the Blue Waters sustained-petascale computing project, which is supported by the National Science Foundation (awards OCI-0725070 and ACI-1238993) and the state of Illinois. Blue Waters is a joint effort of the University of Illinois at Urbana-Champaign and its National Center for Supercomputing Applications. This work is also part of the PRAC “Lattice QCD on Blue Waters”. This research used resources of the National Energy Research Scientific Computing Center (NERSC), a DOE Office of Science User Facility supported by the Office of Science of the U.S. Department of Energy under Contract No. DEAC02-05CH11231. The authors acknowledge the Texas Advanced Computing Center (TACC) at The University of Texas at Austin for providing HPC resources that have contributed to the research results reported within this paper.
Gauge configurations were generated using resources awarded from the U.S. Department of Energy INCITE program at Oak Ridge National Lab, NERSC, the NSF Teragrid at the Texas Advanced Computer Center and the Pittsburgh Supercomputer Center, as well as at Jefferson Lab.
\end{acknowledgments}

\appendix
\section*{Appendices}

\section{Operator Lists}
\label{app:sec:ops}

In tables~\ref{tab:ops} and \ref{tab:ops2} we summarise the operators used in this study.

\begin{table}[htp]
	\centering
	\begin{tabular}{c|c|c|c|c}
		$A_1^+ [000]$ & $A_1 [100]$ & $A_1 [110]$ & $A_1 [111]$ & $A_1 [200]$ \\
		\hline
		$D_{[000]}$ $\pi_{[000]}$  & $D_{[000]}$ $\pi_{[100]}$  & $D_{[000]}$ $\pi_{[110]}$  & $D_{[000]}$ $\pi_{[111]}$  & $D_{[100]}$ $\pi_{[100]}$  \\ 
		$D_{[100]}$ $\pi_{[100]}$  & $D_{[100]}$ $\pi_{[000]}$  & $D_{[100]}$ $\pi_{[100]}$  & $D_{[100]}$ $\pi_{[110]}$  & $D_{[110]}$ $\pi_{[110]}$  \\ 
		$D_{[110]}$ $\pi_{[110]}$  & $D_{[100]}$ $\pi_{[110]}$  & $D_{[110]}$ $\pi_{[000]}$  & $D_{[110]}$ $\pi_{[100]}$  & $D_{[200]}$ $\pi_{[000]}$  \\ 
		$D_{[111]}$ $\pi_{[111]}$  & $D_{[100]}$ $\pi_{[200]}$  & $D_{[110]}$ $\pi_{[110]}$  & $D_{[111]}$ $\pi_{[000]}$  & $D_{[210]}$ $\pi_{[100]}$  \\ 
		$D_{[000]}$ $\eta_{[000]}$  & $D_{[110]}$ $\pi_{[100]}$  & $D_{[111]}$ $\pi_{[100]}$  & $D_{[211]}$ $\pi_{[100]}$  & $D_{[200]}$ $\eta_{[000]}$  \\ 
		$D_{[100]}$ $\eta_{[100]}$  & $D_{[110]}$ $\pi_{[111]}$  & $D_{[210]}$ $\pi_{[100]}$  & ${D^*}_{[110]}$ $\pi_{[100]}$  &  \\ 
		${D_s}_{[000]}$ $\bar{K}_{[000]}$  & $D_{[111]}$ $\pi_{[110]}$  & ${D^*}_{[100]}$ $\pi_{[100]}$  & $D_{[111]}$ $\eta_{[000]}$  &  \\ 
		& $D_{[200]}$ $\pi_{[100]}$  & ${D^*}_{[111]}$ $\pi_{[100]}$  & ${D_s}_{[111]}$ $\bar{K}_{[000]}$  &  \\ 
		& $D_{[210]}$ $\pi_{[110]}$  & $D_{[110]}$ $\eta_{[000]}$  &  &  \\ 
		& $D_{[000]}$ $\eta_{[100]}$  & ${D_s}_{[110]}$ $\bar{K}_{[000]}$  &  &  \\ 
		& $D_{[100]}$ $\eta_{[000]}$  &  &  &  \\ 
		& ${D_s}_{[000]}$ $\bar{K}_{[100]}$  &  &  &  \\ 
		& ${D_s}_{[100]}$ $\bar{K}_{[000]}$  &  &  &  \\ 
		$8\times \bar{\psi} {\bm\Gamma} \psi$ & $18\times\bar{\psi} {\bm\Gamma} \psi$ & $18\times\bar{\psi} {\bm\Gamma} \psi$  & $9\times\bar{\psi} {\bm\Gamma} \psi$ & $16\times\bar{\psi} {\bm\Gamma} \psi$ \\ 
	\end{tabular}
	\caption{Operators used in the variational analyses for irreps featuring $S$-wave $D\pi$ as the lowest subduced partial wave. Subscripts indicate momentum types. $\bm\Gamma$ represents some monomial of $\gamma$ matrices and derivatives.}
	\label{tab:ops}
\end{table}

\begin{table}[htp]
	\centering
	\begin{tabular}{c|c|c|c}
		$T_1^- [000]$ & $E_2 [100]$ & $B_1 [110]$ & $B_2 [110]$ \\
		\hline
		$D_{[100]}$ $\pi_{[100]}$  & $D_{[100]}$ $\pi_{[110]}$  & $D_{[100]}$ $\pi_{[100]}$  & $D_{[100]}$ $\pi_{[111]}$  \\ 
		$D_{[110]}$ $\pi_{[110]}$  & $D_{[110]}$ $\pi_{[100]}$  & $D_{[110]}$ $\pi_{[110]}$  & $D_{[110]}$ $\pi_{[110]}$  \\ 
		${D^*}_{[100]}$ $\pi_{[100]}$  & ${D^*}_{[000]}$ $\pi_{[100]}$  & $D_{[210]}$ $\pi_{[100]}$  & $D_{[111]}$ $\pi_{[100]}$  \\ 
		& ${D^*}_{[100]}$ $\pi_{[000]}$  & ${D^*}_{[100]}$ $\pi_{[100]}$  & ${D^*}_{[000]}$ $\pi_{[110]}$  \\ 
		&  & ${D^*}_{[110]}$ $\pi_{[000]}$  & ${D^*}_{[100]}$ $\pi_{[100]}$ $\{2\}$ \\ 
		&  &  & ${D^*}_{[110]}$ $\pi_{[000]}$  \\ 
		&  &  & ${D^*}_{[111]}$ $\pi_{[100]}$  \\ 
		$6\times \bar{\psi} {\bm\Gamma} \psi$ & $18\times\bar{\psi} {\bm\Gamma} \psi$  & $18\times\bar{\psi} {\bm\Gamma} \psi$ & $20\times\bar{\psi} {\bm\Gamma }\psi$  \\
	\end{tabular}
	\caption{As table~\ref{tab:ops}, but for operators used in irreps without an $S$-wave $D\pi$ subduction. The number in curly parentheses indicates the number of operators of this momentum combination. This arises due to the $D^\star$ appearing in both $[100]A_1$ and $[100]E_2$ that when combined with a pion in $[100]A_2$ results in two linearly independent operators in $[110]B_2$.}
		\label{tab:ops2}
\end{table}

\section{Relation between unitarised chiral amplitudes and the $K$-matrix}
\label{app:sec:chipt}

There have been several applications of unitarised chiral amplitudes to heavy-light systems such as $D\pi$ and $DK$~\cite{Hofmann:2003je,Guo:2008gp,Guo:2009ct,Albaladejo:2016lbb,Guo:2018kno,Guo:2018tjx}. We use the amplitude definitions from a recent example~\cite{Guo:2018tjx}, that considered the coupled-channel $D\pi,D\eta,D_s\bar{K}$ spectra from ref.~\cite{Moir:2016srx}. In this implementation, we only consider the elastic $D\pi$ channel and fit only to the spectra presented above with no other inputs.

Taking the ``loop function'' $G_{\mathrm{DR}}$ as defined in Eq.~14 of ref.~\cite{Guo:2018tjx} and the Chew-Mandelstam function $I(s)$ as defined in Appendix B of ref.~\cite{Wilson:2014cna}, it is straightforward to show that they differ only by normalisation and terms independent of $s$. If the subtraction point of the Chew-Mandelstam function is chosen as threshold and $I(s_{\mathrm{thr.}})=0$ then the relation is,
\begin{align}
16\pi\: G_{\mathrm{DR}}(s,m_1,m_2)=I(s) + \frac{\alpha(\mu)}{\pi} + \frac{2}{\pi}\left(\frac{m_2}{m_1+m_2}\log\frac{m_2}{m_1} + \log\frac{m_1}{\mu}\right)
\label{eq_grel}
\end{align} 
where $m_1=m_\pi$ and $m_2=m_D$ are the scattering particle masses, $\alpha(\mu)$ and $\mu$ appear in $G_{\mathrm{DR}}$, where $\mu$ is an energy scale taken to be 1 GeV and $\alpha(\mu)\approx-1.8$~\cite{Oller:2000fj}.

If $\mathcal{V}_{J=0}$ is the $S$-wave projected elastic $D\pi$ scattering amplitude, then in the definitions used throughout this paper this can be written as,
\begin{align}
K^{-1}(s)&=\left(-\frac{1}{16\pi}\mathcal{V}_{J=0}\right)^{-1}+\frac{\alpha(\mu)}{\pi} + \frac{2}{\pi}\left(\frac{m_2}{m_1+m_2}\log\frac{m_2}{m_1} + \log\frac{m_1}{\mu}\right)\;.
\label{eq_K_chipt_app}
\end{align}
Using only the leading order expression for $\mathcal{V}_{J=0}$, 
\begin{align}
\mathcal{V}_{J=0} &= \frac{C_{\mathrm{LO}}}{8sF^2}\left(3s^2-2s(m_1^2+m_2^2)-(m_2^2-m_1^2)^2\right)
\end{align}
where $F\approx f_\pi$, and $C_{\mathrm{LO}}=-2$ for $I=1/2$ $D\pi$ scattering~\cite{Guo:2018tjx}. Equation~\ref{eq_K_chipt_app} can then be written as a ratio of polynomials up to $\mathcal{O}(s^2)$ as in Eq.~\ref{eq_K_ratio}. For simplicity, we do not consider higher order terms and allow $F$ and $\alpha(\mu)$ to float such that the spectra are well described. We have verified that the next-to-leading order term does not make a large change to the amplitude using typical values for the next-to-leading order Wilson coefficients given in the literature, when applied at $m_\pi\approx 239$ MeV.

\section{Simultaneously fitting $D\pi$ with $J^P=0^+,1^-$ and $D^\star\pi$ with $J^P=1^+$}
\label{app:sec:fullfit}

\begin{figure}[tb]
	\centering
	\graphicspath{{plots/}}
	\input{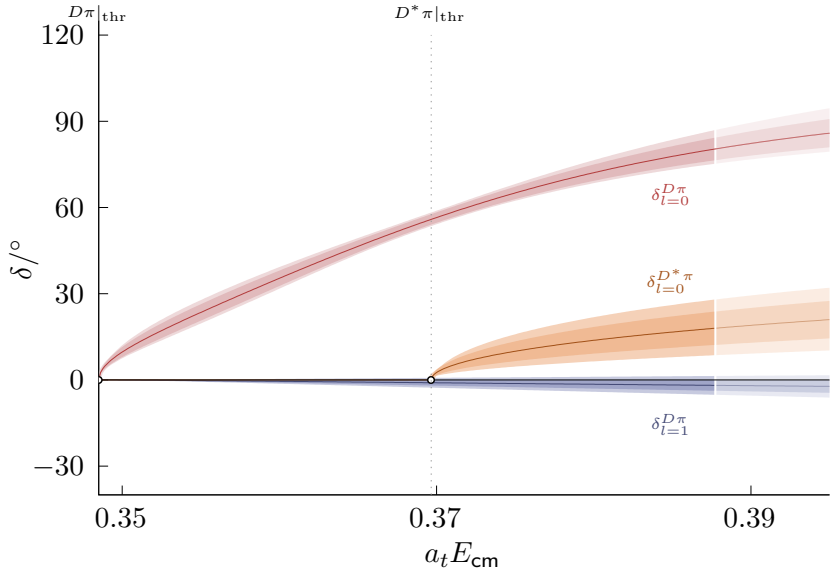}
	\caption{As Fig.~\ref{P_phase_shift}, but for phase shifts of $D\pi$ in $S$-wave (red) and $P$-wave (blue), and $D^\star\pi$ in $S$-wave (orange) corresponding to Eq.~\ref{eq_SPmPp}.}
	\label{fit:fullfit_phase_shift}
\end{figure}

As an additional check, we also perform a simultaneous fit to all the black points shown in Figs.~\ref{fig:spec1} and \ref{fig:spec2}, except $[000]E^+$. We use $D \pi$ amplitudes with $J^P=0^+,1^-$, and the $J^P=1^+$ $D^* \pi$ in a relative $S$-wave, neglecting any $D^\star\pi$ $D$-wave dynamical mixing. The $S$ and $P$-waves are parameterised as Eq.~\ref{eq_ref_amp}, and the $D^* \pi$ $1^+$ wave is parameterised as in Eq.~\ref{eq_PmPp}. After minimising the $\chi^2$ to best describe the spectra, we obtain
\begin{equation*}
\begin{aligned}[t]
\begin{matrix}
m &= &(0.403  \pm 0.020 \pm 0.004) \cdot a_t^{-1} \\
g &= &(0.43  \pm 0.19 \pm 0.02) \cdot a_t^{-1} \\
\gamma^{(0)} &= &(-2.2  \pm 2.8 \pm 1.8) \\
m_1 &= &(0.33024 \pm 0.00016 \pm 0.00002) \cdot a_t^{-1} \\
g_1 &= &(0.54  \pm 0.72 \pm 0.00) \cdot a_t^{-1} \\
\gamma^{(0)} [D^* \pi] &= &(1.67  \pm 0.75 \pm 0.11)
\end{matrix}
\end{aligned}
\qquad
\begin{aligned}[t]
\begin{bmatrix}
1.00 & 0.98 & -0.92 & 0.37 & 0.53 & -0.56 \\
& 1.00 & -0.97 & 0.35 & 0.59 & -0.64 \\
& & 1.00 & -0.32 & -0.66 & 0.74 \\
& & & 1.00 & 0.30 & -0.27 \\
& & & & 1.00 & -0.72 \\
& & & & & 1.00
\end{bmatrix}
\end{aligned}
\end{equation*}
\begin{equation}
\chi^2/N_{\text{dof}} = 21.64 / (29 - 6) =  0.94\,.
\label{eq_SPmPp}
\end{equation}
The corresponding phase shifts are shown in Fig. \ref{fit:fullfit_phase_shift}, where it can be seen that the $D\pi$ phase shifts are very similar to those shown in Fig. \ref{SP_phase_shift}. A positive phase shift is found in the $D^\ast\pi$, similar to that shown in Fig.~\ref{P_phase_shift}.

\section{$P$-wave amplitude plotted as $k^3\cot\delta_1$}
\label{app:kcot}

The irreps that are sensitive to the $J^P=1^-$ amplitude all contain a level far below threshold. In the reference amplitude Eq.~\ref{eq_ref_amp}, we found that these energy levels could be described by a $K$-matrix with a pole term, that was found to correspond to a deeply bound state, as explained in Sec.~\ref{sec:poles}. The $K$-matrix pole coupling term was found to be consistent with zero in a number of fits, and no influence was found in the amplitudes above $D\pi$ threshold, as can be seen in Figs.~\ref{P_phase_shift}, ~\ref{SP_phase_shift} and \ref{fit:fullfit_phase_shift} where the $P$-wave phase shifts are all small or consistent with zero.

\begin{figure}	\centering
	\graphicspath{{plots/}}
	\includegraphics[width=0.66\textwidth]{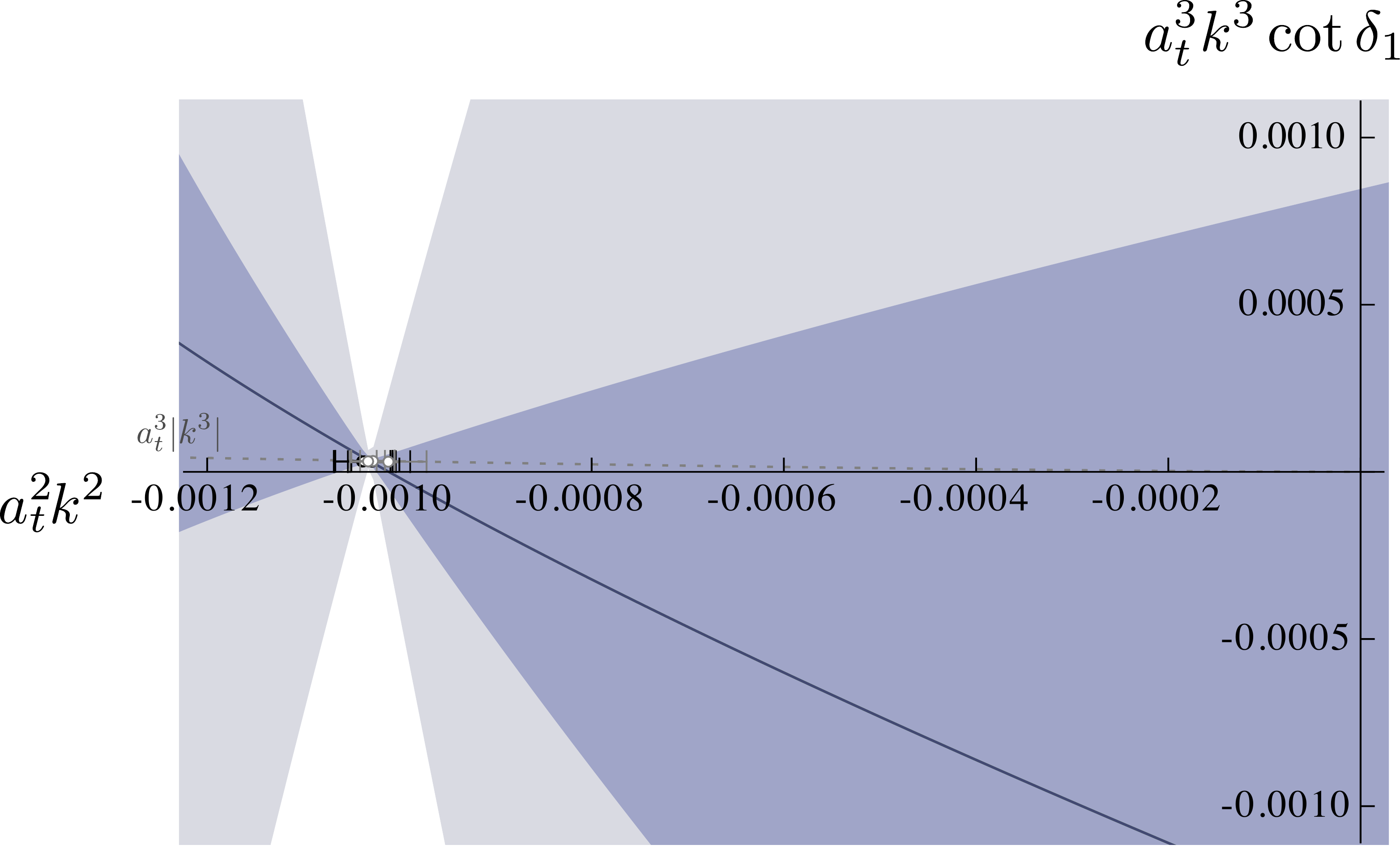}
	\caption{The $P$-wave part of the reference amplitude Eq.~\ref{eq_ref_amp} plotted as $k^3\cot\delta_1$. The grey points are from the moving frame $A_1$ irreps where $S$-wave also appears, the black points are from $[000]T_1^-$ and the other moving frame irreps. The meaning of the bands is as in Fig.~\ref{SP_phase_shift}. The dotted line shows $a_t^3|k^3|$.}
	\label{fig_P_ere}
\end{figure}

The $P$-wave amplitude may be plotted as $k^3\cot\delta_1$ analogously to as was done for the $S$-wave plotted in Figs.~\ref{fig_860_ere} and \ref{fig:kcot}. This is shown in Fig.~\ref{fig_P_ere} for the reference amplitude given in Eq.~\ref{eq_ref_amp}. The amplitude is constrained below threshold at the points shown, and also above threshold by energy levels where $P$-wave is leading from $[000]T_1^-$ and as a subleading wave in the moving frame $A_1$ irreps. The result is an amplitude that has a pole far below threshold and produces a small phase shift above threshold.

\bibliography{biblio}
\bibliographystyle{JHEP}

\end{document}